\RequirePackage{fix-cm}
\documentclass[natbib,smallextended,twocolumn]{svjour3}          % twocolumn
\smartqed  % flush right qed marks, e.g. at end of proof
\usepackage{graphicx}
\usepackage{amsmath}	% Advanced maths commands
\usepackage{amssymb}	% Extra maths symbols
\usepackage{array}
\usepackage{subfig}
\usepackage{newtxtext,newtxmath}
\usepackage[T1]{fontenc}
\usepackage{hyperref}       % hyperlinks
\usepackage{lineno}
\usepackage{xcolor}
\usepackage{color}
\definecolor{myorange}{RGB}{245,156,74}
\definecolor{darkblue}{rgb}{0.,0.,0.4}
\definecolor{darkred}{rgb}{0.5,0.,0.}
\hypersetup{colorlinks=True,urlcolor=darkred, citecolor=cyan, linkcolor=blue}

\usepackage{cancel}

\journalname{Experimental Astronomy}
\begin{document}

\title{MeerCRAB: MeerLICHT Classification of Real and Bogus Transients using Deep Learning}

\titlerunning{MeerCRAB: MeerLICHT Classification of Real and Bogus Transients using Deep Learning}        % if too long for running head

\author{Zafiirah Hosenie$^{1}$$^{*}$ \and
Steven Bloemen$^{2}$ \and
Paul Groot$^{2,3,4}$ \and
Robert Lyon$^{5}$ \and
Bart Scheers$^{6}$ \and
Benjamin Stappers$^{1}$ \and
Fiorenzo Stoppa$^{2}$ \and
Paul Vreeswijk$^{2}$ \and
Simon De Wet$^{3}$ \and
Marc Klein Wolt$^{2}$ \and
Elmar K\"ording$^{2}$ \and 
Vanessa McBride$^{3,7}$ \and
Rudolf Le Poole$^{8}$ \and
Kerry Paterson$^{9}$ \and
Dani\"elle L. A. Pieterse$^{2}$ \and
Patrick Woudt$^{3}$
}

\institute{\at $^{1}$Jodrell Bank Centre for Astrophysics, Department of Physics and Astronomy, The University of Manchester, Manchester M13 9PL, UK. \\
              \email{\href{zafiirah.hosenie@gmail.com}{zafiirah.hosenie@gmail.com}}
           \and
           \at $^{2}$Department of Astrophysics/IMAPP, Radboud University, P.O. 9010,6500 GL, Nijmegen, The Netherlands. \\
           \at $^{3}$Inter-University Institute for Data Intensive Astronomy \& Department of Astronomy, University of Cape Town, Private Bag X3, Rondebosch 7701, South Africa.\\ 
           \at $^{4}$South African Astronomical Observatory, P.O. Box 9, 7935 Observatory, South Africa.\\
           \at $^{5}$Department of Computer Science, Edge Hill University, Ormskirk Lancashire L39 4QP, UK.\\
           \at $^{6}$Dataspex B.V., c/o Centrum Wiskunde \& Informatica, PO Box 94079, 1090 GB Amsterdam, The Netherlands.\\
           \at $^{7}$IAU-Office For Astronomy for Development, P.O. Box 9, 7935 Observatory, South Africa.\\ 
           \at $^{8}$Leiden Observatory, Leiden University, P.O. Box 9513, NL-2300 RA Leiden, The Netherlands.\\
           \at $^{9}$Center for Interdisciplinary Exploration and Research in Astrophysics (CIERA) and Department of Physics and Astronomy, Northwestern University, Evanston, IL 60208, USA.\\ 
           }

\authorrunning{Z.Hosenie et al. 2021} % if too long for running head

% \institute{
%               \email{\href{zafiirah.hosenie@gmail.com}{zafiirah.hosenie@gmail.com}}
% }

\date{Accepted: 22 April 2021}
%\date{Received: xxx 2021 / Accepted: xxx 2021}

\label{firstpage}
\maketitle

\begin{abstract}
Astronomers require efficient automated detection and classification pipelines when conducting large-scale surveys of the (optical) sky for variable and transient sources. Such pipelines are fundamentally important, as they permit rapid follow-up and analysis of those detections most likely to be of scientific value. We therefore present a deep learning pipeline based on the convolutional neural network architecture called \texttt{MeerCRAB}. It is designed to filter out the so called ``bogus'' detections from true astrophysical sources in the transient detection pipeline of the MeerLICHT telescope. Optical candidates are described using a variety of 2D images and numerical features extracted from those images. The relationship between the input images and the target classes is unclear, since the ground truth is poorly defined and often the subject of debate. This makes it difficult to determine which source of information should be used to train a classification algorithm. We therefore used two methods for labelling our data (i) thresholding and (ii) latent class model approaches. We deployed variants of \texttt{MeerCRAB} that employed different network architectures trained using different combinations of input images and training set choices, based on classification labels provided by volunteers. The deepest network worked best with an accuracy of 99.5\% and Matthews correlation coefficient (MCC) value of 0.989. The best model was integrated to the MeerLICHT transient vetting pipeline, enabling the accurate and efficient classification of detected transients that allows researchers to select the most promising candidates for their research goals.
\end{abstract}


%%%%%%%%%%%%%%%%%%%%%%%%%%%%%%%%%%%%%%%%%%%%%%%%%%

%%%%%%%%%%%%%%%%% BODY OF PAPER %%%%%%%%%%%%%%%%%%

\section{Introduction}

Contemporary large-scale optical surveys such as Skymapper \citep{2007pasa...24....1k}, the Palomar Transient Factory (PTF, \citealt{2009PASP..121.1334R}), the Catalina Real-time Transient Survey (CRTS, \citealt{2009apj...696..870d}), the Panoramic Survey Telescope and Rapid Response System (Pan-STARRS1, \citealt{2010SPIE.7733E..0EK}), the All-Sky Automated Survey for SuperNova (ASASSN, \citealt{2014apj...788...48s}), Gaia \citep{2016A&A...595A...2G}, the MeerLICHT telescope \citep{2016SPIE.9906E..64B, 2019IAUS..339..203P} and the Zwicky Transient Factory (ZTF, \citealt{2019PASP..131a8002B}) are generating a plethora of transient events originating from a wide range of sources. These instruments enable us to observe and explore changes in millions of sources/candidates, thus unlocking new opportunities for interpreting and understanding large families of sources. 

The MeerLICHT facility provides a 2.7 square degree field-of-view at a pixel scale of 0.56"/pixel \citep{2016SPIE.9906E..64B} that maximises the volume of astrophysical candidates with brightnesses appropriate for spectroscopic follow-up using current large-aperture optical facilities. More details regarding the survey can be found in \citet{2016SPIE.9906E..64B}. Both MeerLICHT and the BlackGEM array \citep{2019NatAs...3.1160G} (that is currently being installed at the La Silla Observatory in Chile) will yield about 500 observations per night, per telescope, thus generating hundreds of candidate alerts every clear night that could be spectroscopically followed up. BlackGEM's main focus is on the detection of optical counterparts to gravitational wave events and MeerLICHT is used to co-observe the sky as seen with the MeerKAT radio array \citep{2016mks..confE...1J}. MeerLICHT and BlackGEM are technically identical with MeerLICHT being the prototype for the BlackGEM array.

Transients and variables are sources that vary on all timescales (from milliseconds up to years) and they vary significantly from a reference image - either an increase or decrease in brightness. Transients include phenomena such as supernovae, gamma-ray bursts, tidal disruption events and flare stars, to name a few. A successful transient follow-up program enables the creation of a large database of transient and variable sources. Such large databases are important for future analyses of data collected during upcoming photometric surveys such as those conducted at the Vera C. Rubin observatory (LSST; \citealt{2009arXiv0912.0201L}). While we possess a reasonable understanding of many transient sources, achieved via consideration of their spectra, the main goal of surveys undertaken with MeerLICHT is to find and select the subset of sources that are not well understood. This will help us  to increase our knowledge of the different families of transients and variable stars. Secondly, given that transients are rapidly fading sources due to their often destructive nature, MeerLICHT aims to identify transients rapidly, as they are only visible for a limited amount of time for follow-up. 

\begin{figure}
\centering
\includegraphics[width=0.5\textwidth]{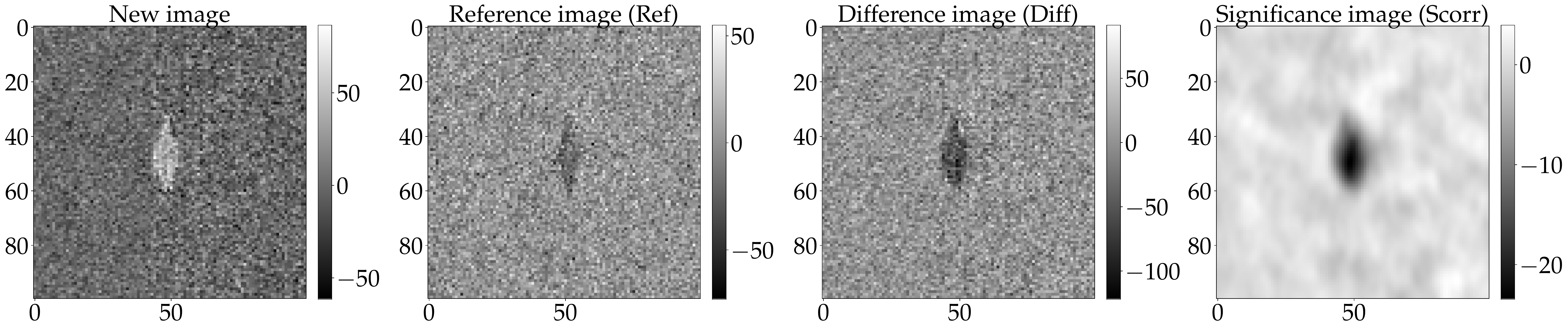}\\
\subfloat[Bogus examples.]{\includegraphics[width=0.5\textwidth]{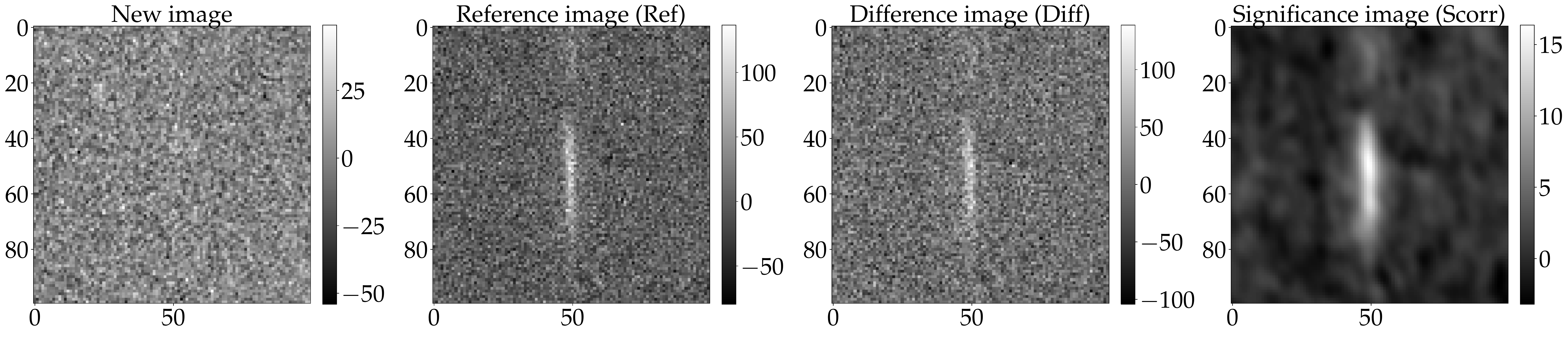}}\\
\includegraphics[width=0.5\textwidth]{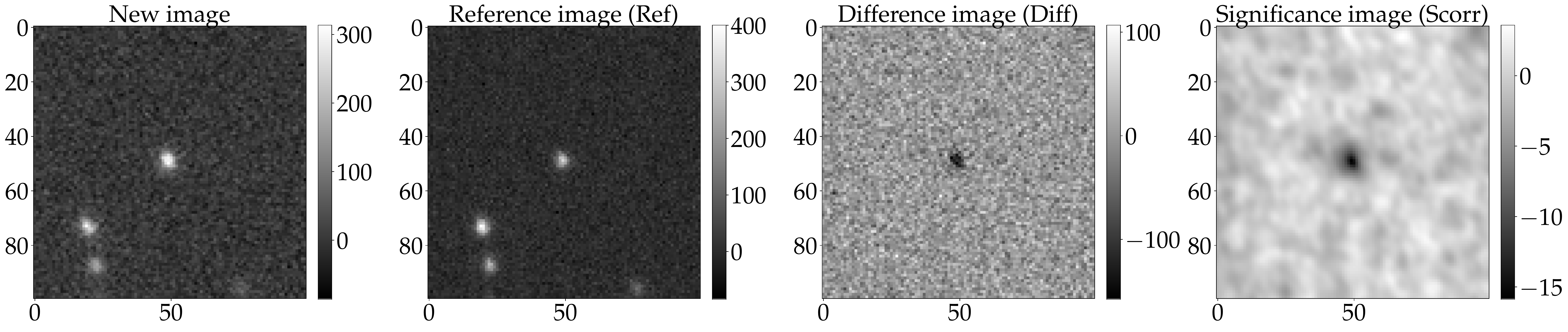}\\
\subfloat[Real examples.]{\includegraphics[width=0.5\textwidth]{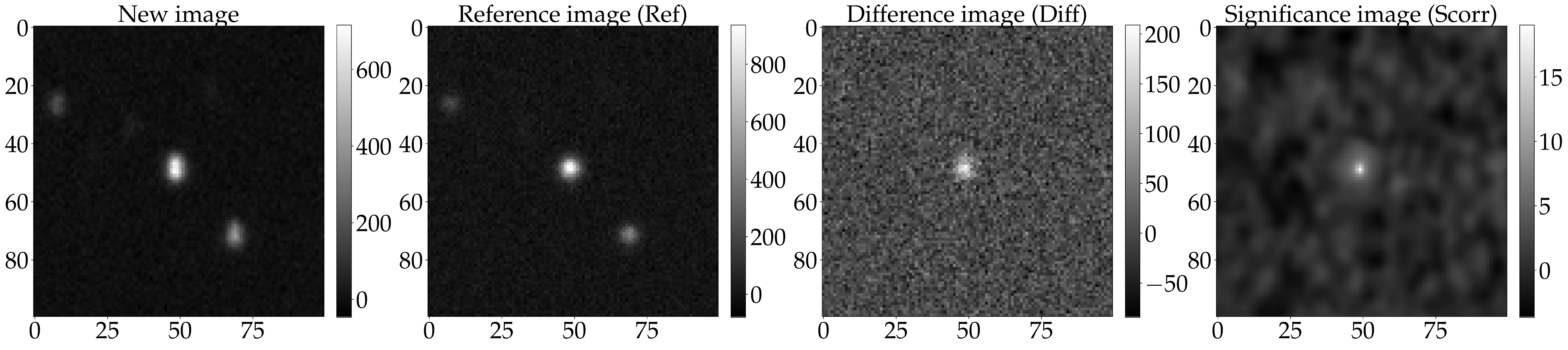}}\\
\caption{\label{fig:real-bogus-images}Examples of bogus and real transients in the MeerLICHT database. Each column represents the new (N), reference (R), difference (D) and significance (S) images and the rows are the different fields.}
\end{figure}

\begin{figure*}
	\centering
	\includegraphics[width=0.8\textwidth]{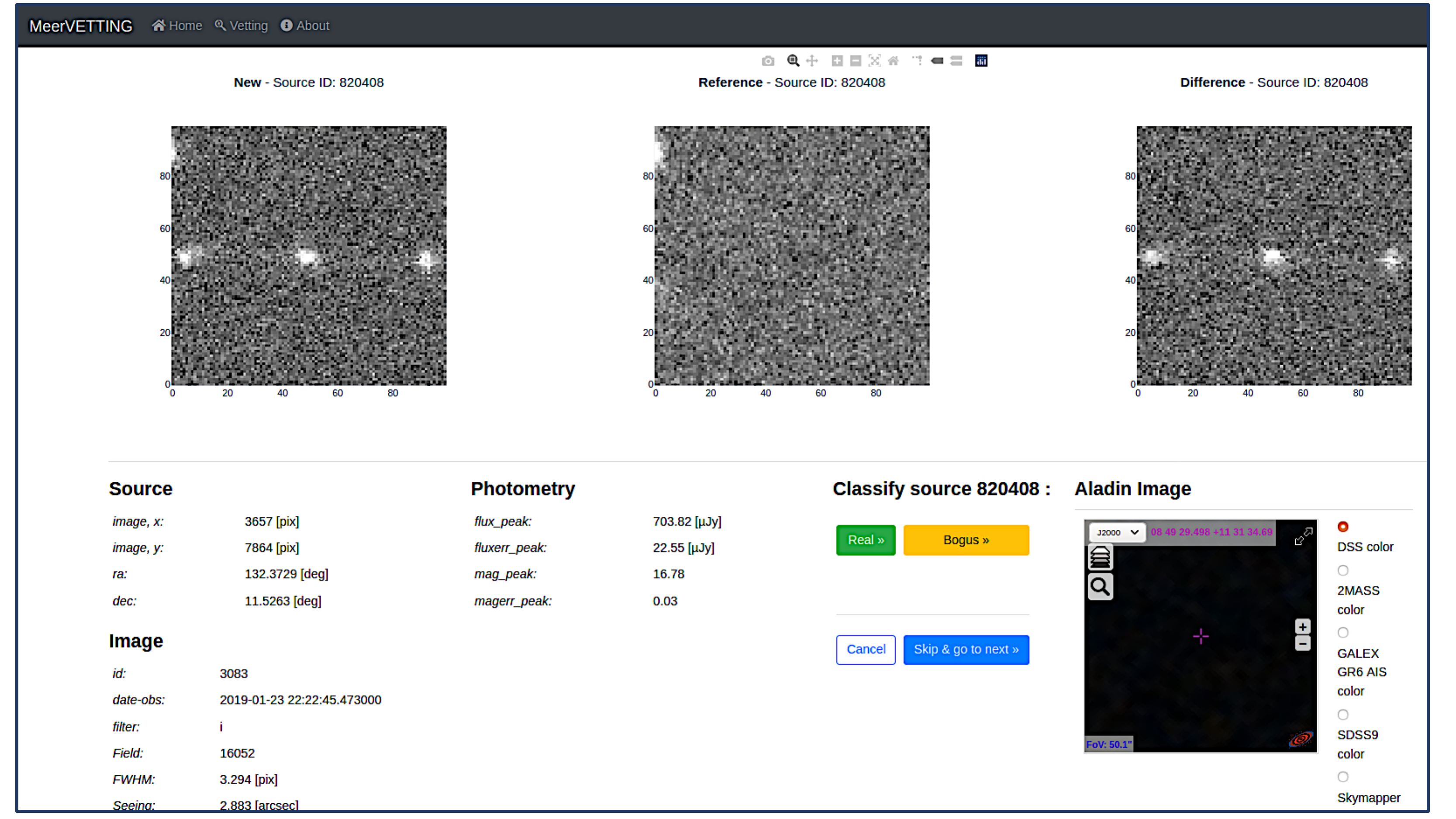}
    \caption{\texttt{MeerVETTING} web-interface used to label MeerLICHT candidates as either real, bogus or skip confused candidates. Vetters are provided with three images (new, reference, difference) that are cut-outs of the science images.}
    \label{fig:meervetting-web-interface}
\end{figure*}

In order to have an early and rapid characterisation of these sources, it is fundamentally important to automate several steps within a transient detection pipeline, including the separation of transients/astrophysical events from ``bogus'' detections, which has become a bottle-neck in fast detection pipelines. So called ``bogus'' detections can occur as a result of saturated sources, convolution problems, defects in the detector, atmospheric dispersion, unmodeled differences at the subtraction stage and cosmic rays passing through the detector, amongst other things. 

Most surveys use three images for transient event detection and extraction: (i) an early observation of the relevant sky (also known as the \textit{template}/\textit{reference} image), (ii) a calibrated recent image (\textit{New}/\textit{Science} image), (iii) the \textit{difference} image which is formed by subtracting the \textit{reference} image from the \textit{new}/\textit{science} image. Using the difference image, one can, in principle, effectively detect transients, however, in many cases, the subtracted image contains bogus sources. 

Therefore to be successful, surveys require an automated way to distinguish between real and bogus candidates. To address this challenging task, most of the time-domain surveys mentioned previously have adopted machine learning (ML) algorithms to perform real-bogus classification. Convolutional neural networks (CNNs, \citealt{Lecun99objectrecognition}) have been used in the image domain as feature extractors for automatic vetting algorithms, for example, during the Skymapper Survey \citep{2017MNRAS.472.3101G}, the High cadence Transient Survey (HiTS, \citealt{2017ApJ...836...97C}) and the ZTF \citep{2019PASP..131a8002B} similarly utilized deep learning techniques. Other ML techniques such as Random Forest (RF) and \textit{k}-Nearest Neighbour (\textit{k}-NN) classification approaches have been employed to classify light curve transients from CRTS \citep{2011ApJ...733...10R, 2020MNRAS.493.6050H, 2019MNRAS.488.4858H}. 

%\ZH{When using ML based automated classification systems, we should not use models trained on data acquired at one telescope, to make predictions upon data acquired by another. Doing so constitutes a violation of the i.i.d principle, which ultimately limits classification performance and consistency. The per- formance of an ML system is thus entirely dependent upon the quality and distributional properties of the input data it is given. For a system to perform well for a given task, it must be built using data that is distributionally similar to the data it must process in practice.} 

The classification task in these surveys is usually separated into two distinct steps. Firstly, bogus candidates are filtered out from real sources immediately after acquiring data, that is, the classification between real and bogus. The second stage involves assigning an astrophysical category/class label to each detected transient based on its spectroscopic or photometric information (e.g. \citealt{2019PASP..131k8002M}). In this paper, we focus on the automation of the first stage, that is, the classification of sources as either Real or Bogus using deep learning methods developed for the MeerLICHT facility.

We note that when using ML based automated classification systems, we should not use models trained on data acquired at one telescope, to make predictions upon data acquired by another. Doing so constitutes a violation of the i.i.d principle, which ultimately limits classification performance and consistency. In addition, labelling mistakes are often even more costly - especially on rare sub-classes of transient phenomena. The performance of an ML system is thus entirely dependent upon the quality and distributional properties of the input data and the associated labels it is given. For a system to perform well for a given task, it must be built using data and labels that are distributionally similar to the data it must process in practice. In this work, we present two labelling strategies to label our data (i) \textit{thresholding} which removes noisy labelling and (ii) the \textit{Latent class model}, $\textrm{L}_{lcm}$ \citep{formann1984latent} which incorporates labelling uncertainty in our model. Afterwards, we constructed three models based on CNNs to build a new robust system that separates real candidates from their bogus counterparts for the MeerLICHT-transient search pipeline. In \S\ref{sec:MeerLICHT_facility} we provide an overview of the MeerLICHT telescope and we detail the data used for training and testing the \texttt{MeerCRAB} algorithms. In \S\ref{sec:MeerCRAB-pipeline}, the methods, network set-up and architectures are described. Results and experimental set-up are detailed in \S\ref{sec:results_and_analysis}, followed by our main conclusions in \S\ref{sec:conclusions}.

\section{The MeerLICHT facility}\label{sec:MeerLICHT_facility}

MeerLICHT is an optical wide-field telescope that is operated robotically. The telescope is located at the Sutherland station of the South African Astronomical Observatory (SAAO). It consists of a 65 cm primary mirror and provides a 2.7 square degree field-of-view  at a pixel scale of 0.56"/pixel \citep{2016SPIE.9906E..64B}. MeerLICHT will co-observe with the MeerKAT radio telescope on the same field. The combination of an optical and a radio telescope will enable the study of fast transient phenomena using simultaneous observations in two very distinct parts of the electromagnetic spectrum, whilst eliminating the delay introduced by triggering optical follow-up after the detection of a radio event.

\begin{figure}
	\includegraphics[width=0.5\textwidth]{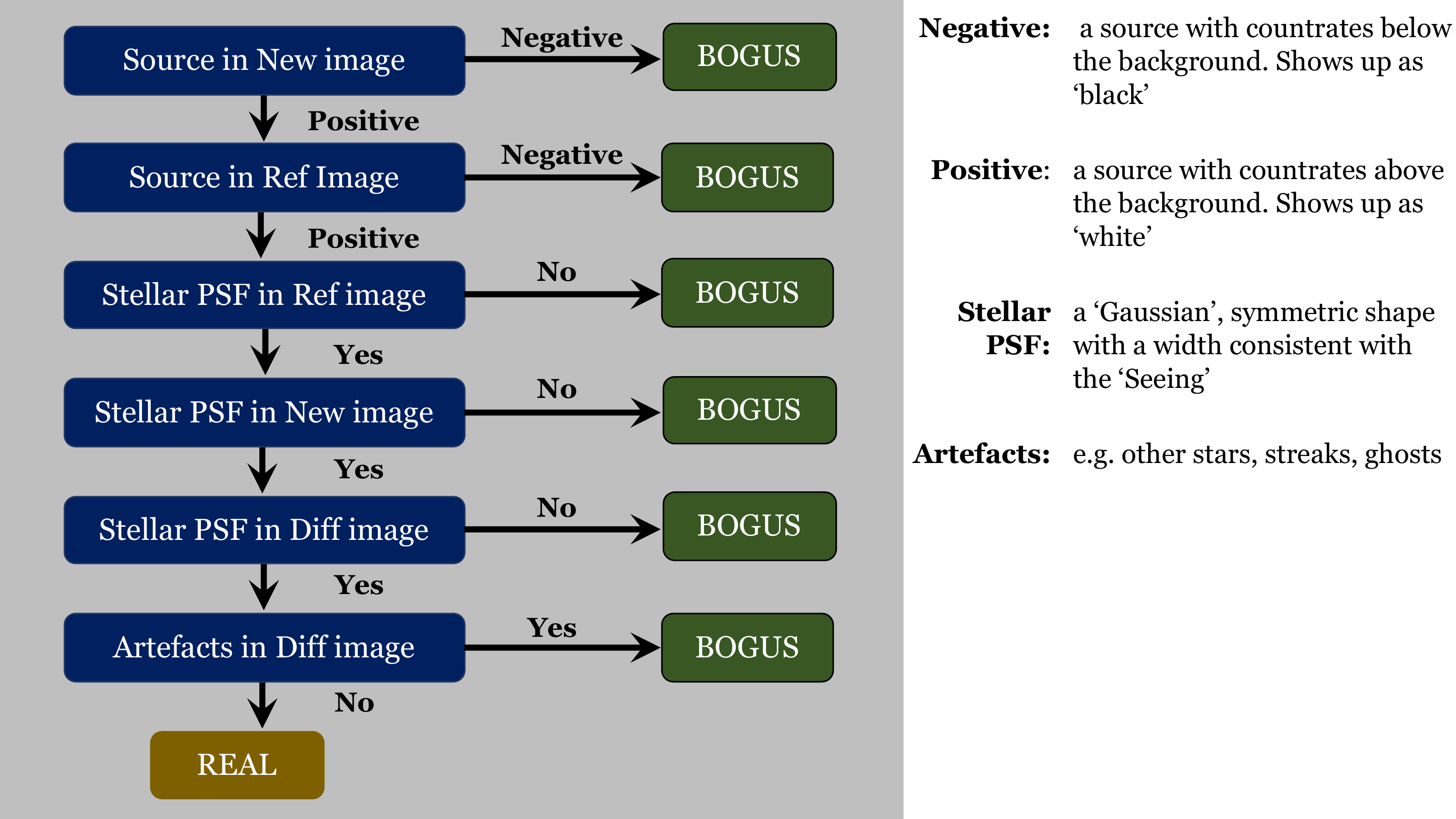}
    \caption{Decision tree characterising real and bogus candidates. Vetters used this guide to label each candidate and to construct a large training set for \texttt{MeerCRAB}.}
    \label{fig:vetting_decision_tree}
\end{figure}

MeerLICHT (and also BlackGEM) images are processed by the BlackBOX package\footnote{see \href{https://github.com/pmvreeswijk/BlackBOX}{https://github.com/pmvreeswijk/BlackBOX} and \href{https://github.com/pmvreeswijk/ZOGY}{https://github.com/pmvreeswijk/ZOGY}} to produce image products and catalogs of all objects detected as well as transient candidates resulting from optimal image subtraction. The raw MeerLICHT images are automatically transferred from SAAO to the Inter-university Institute for Data Intensive Astronomy (IDIA\footnote{see \href{https://www.idia.ac.za/}{https://www.idia.ac.za/}}) in Cape Town, South Africa, and processed by BlackBOX.

First, the images are gain and overscan-corrected, and flatfielded using a set of twilight flats. Cosmic rays and satellite trails are detected using the astroscrappy\footnote{see \href{https://github.com/astropy/astroscrappy}{https://github.com/astropy/astroscrappy}} implementation of LA Cosmic \citep{2001PASP..113.1420V} and STSDAS satdet\footnote{see \href{https://acstools.readthedocs.io/en/latest/satdet.html}{https://acstools.readthedocs.io/en/latest/satdet.html}}  modules, respectively. Subsequently, the following steps are performed: object detection using Source-Extractor \citep{1996A&AS..117..393B}, astrometric calibration using Astrometry.net \citep{2010AJ....139.1782L}, estimation of the Point Spread Function (PSF) as a function of position using PSFEx \citep{2011ASPC..442..435B} and photometric calibration. The latter is done using a custom-built catalog of calibration stars in the MeerLICHT photometric system based on fitting stellar spectral templates to Gaia, SDSS, PanSTARRS, SkyMapper, 2MASS and GALEX photometry.

Finally, optimal image subtraction is performed, comparing the new image with a reference image, closely following the prescriptions of Zackay, Ofek \& Gal-Yam, a.k.a. ZOGY \citep{2016ApJ...830...27Z}. To allow for the PSF to vary across the image, the full MeerLICHT images are divided into 8 by 8 subimages, on which the ZOGY calculations are applied separately, before inserting the subimages back into a full image. The following images are produced: a difference ($D$) image (see Eq. 13 in the ZOGY paper) and a statistics (also known as significance / $Scorr$, $S$) image (see Eqs. 16 and 17 in the ZOGY paper) providing the probability of a transient being present at a particular position. The $Scorr$ image, $S$, is normalized by the Poisson noise of the input images and the error resulting from the astrometric uncertainty when remapping the reference image to the new image frame; this leads to the $Scorr$ image (see Eq. 25 in the ZOGY paper), which we also refer to as the significance image. The unit of this $Scorr$ image is standard deviations (sigma) and transients above an adopted significance threshold (we used $Scorr$$\geq$12 for the data presented in this paper) are normally selected on the basis that they are potentially significant. In practice, many significant but artificial transients are present in collected data due to cosmic rays, saturated stars, bad pixel regions or other image artefacts; many of these can be filtered out with some basic constraints applied to the size and shape, but for each image, tens of transient candidates remain where we only expect a few astrophysical transients per image.

The MeerLICHT/BlackGEM database ingests the transient catalogs produced by BlackBOX, including 1$\times$1 arcminute thumbnail cut-outs around each transient of the new ($N$), the reference ($R$), the difference ($D$) and significance / $Scorr$ ($S$) image as shown in Figure \ref{fig:real-bogus-images}. The process of creating a training set for \texttt{MeerCRAB} is detailed in the next section.

\subsection{\textbf{MeerLICHT dataset and data labelling}}\label{subsec:data-labelling}
Our goal is to automate the separation of real candidates from bogus objects for the MeerLICHT transient detection pipeline. The main challenge faced when building a supervised automated system is that we need to construct a large labelled data set that can be used to train a ML model. In addition, the data set needs to be representative, that is, we should have a fairly balanced number of real and bogus candidates. If the latter are unavailable, ML algorithms built from such unrepresentative data tend to be biased towards the majority class (e.g. \citealt{2019MNRAS.488.4858H}).

We therefore construct a large representative training dataset for the Real-Bogus challenge by manually vetting a selection of transients, using a web-interface, known as \texttt{MeerVETTING} as shown in Figure \ref{fig:meervetting-web-interface}. Using the MeerLICHT database, a set of 5000 transient candidates were randomly selected from MeerLICHT data taken between 2017 and 2020. A team of 10 people ("vetters") were presented with three 100$\times$100 pixels images during vetting, i.e. the new (N), reference (R), and difference (D). The properties for real and bogus candidates were defined as a phenomenological distinction based solely on the MeerLICHT data, not an astrophysical distinction. In the context of MeerLICHT data, `positive' implies positive pixel values and `negative' points to negative pixel values located at the centre of the 100$\times$100 images.
\begin{figure}
	\centering
	\includegraphics[width=0.5\textwidth]{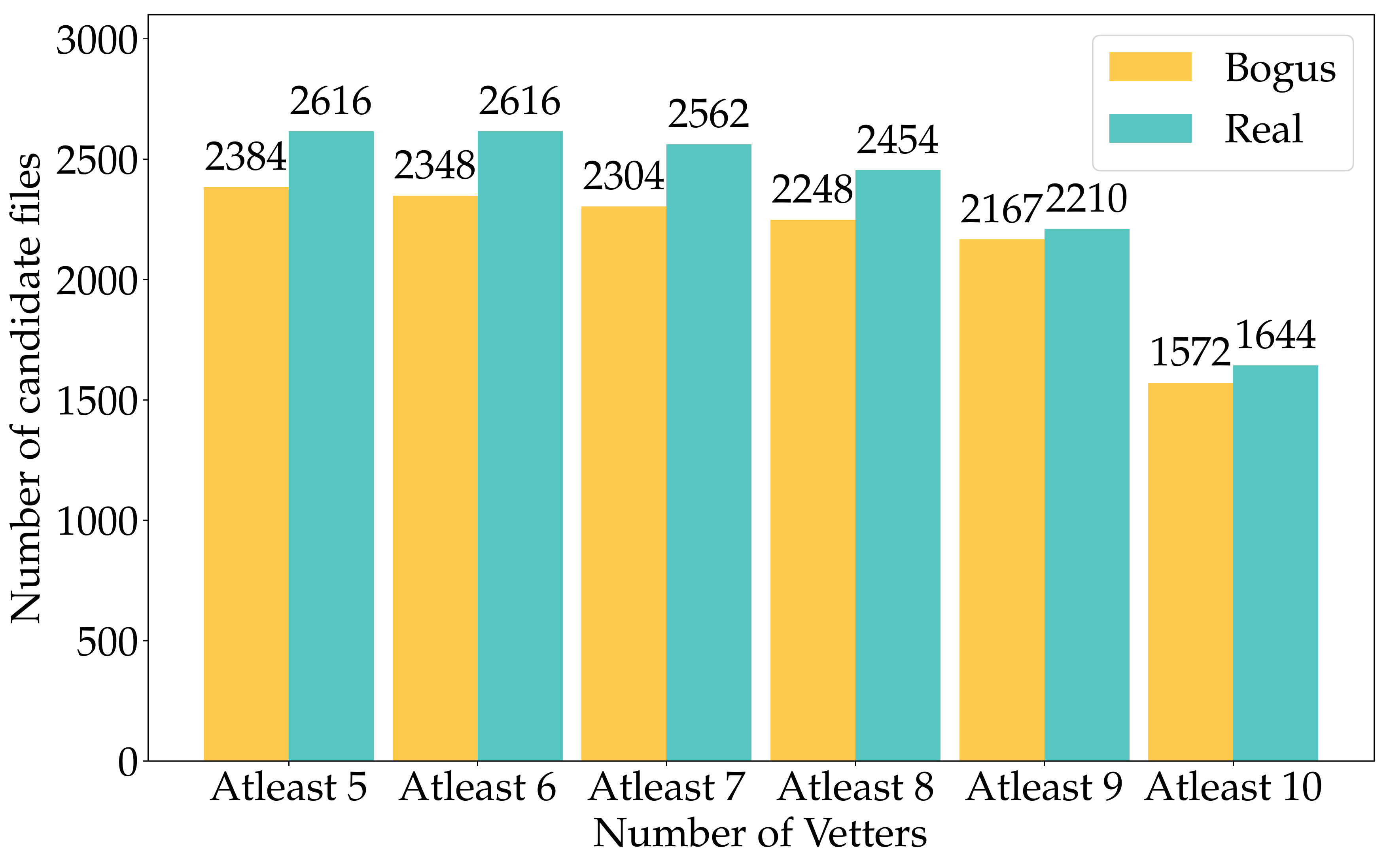}
    \caption{Thresholding method used to analyse the labelling provided by vetters. The x-axis represents criteria we applied on the labelling. For \textit{Atleast 9 (T9)}, this indicates that 9 out of the 10 vetters have agreed on the labelling. Similar strategy applies for \textit{Atleast 5 (T5)} to \textit{Atleast 10 (T10)}. The y-axis is the number of candidates with a given label, either real or bogus.}
    \label{fig:Consistency_in_labelling_5000}
\end{figure}

\begin{itemize}
	\item{Real is any source that is of astrophysical origin, and variable in time and/or position. A real source therefore}
	\begin{enumerate}
		\item{has a shape that reflects a point-source. Most MeerLICHT data is taken in decent focus conditions, so it implies that the source is round, and has a (visual) extent of $\sim$ 5-10 pixels,}
		\item{is positive in either the new or the reference image,}
		\item{can be variable in both directions, e.g. fading or brightening between the new and reference image, and is therefore positive or negative in the difference image and the significance image,}
		\item{can (dis)appear between the new and the reference image. This means that in one of the two images there is no source at all, and in the other there is a clear point-source.}
	\end{enumerate}
	\item{Bogus is any source that is not of astrophysical origin. A bogus source therefore generally has}
	\begin{enumerate}
		\item{a shape that is not a point-source: not round, not `Gaussian`, with a size $\lesssim$ 5 pixels or $\gtrsim$ 10 pixels,}
		\item{is negative in the new image,}
		\item{is positive in the new image but negative in the reference image.}
	\end{enumerate}
\end{itemize}

Before using the \texttt{MeerVETTING} web-interface, vetters were provided with a visual guide of the various properties of real and bogus candidates. By providing vetters with a guide, we in principle create better annotators, who should produce better labels which in turn should yield improved ML models \-- as long as the guide itself is not inherently biased in some way. The above characteristics are summarised in a decision-tree as shown in Figure \ref{fig:vetting_decision_tree}. Using the \texttt{MeerVETTING} web-interface, vetters based their decision to manually vet a source as either Real or Bogus by following the decision-tree.

Despite these guidelines, the decision is still subjective whilst there remain boundary cases that are hard to label. Therefore large training datasets will almost always contain examples with inaccurate labels. We test the performance of the \texttt{MeerCRAB} models by (i) removing confused candidates (noisy labels) using a \textit{thresholding} method, and (ii) including the entire dataset with labels based on the latent class model, $\textrm{L}_{lcm}$. In the following sections, we provide a brief discussion of the two methods.

\begin{figure}
	\centering
	\includegraphics[width=0.5\textwidth]{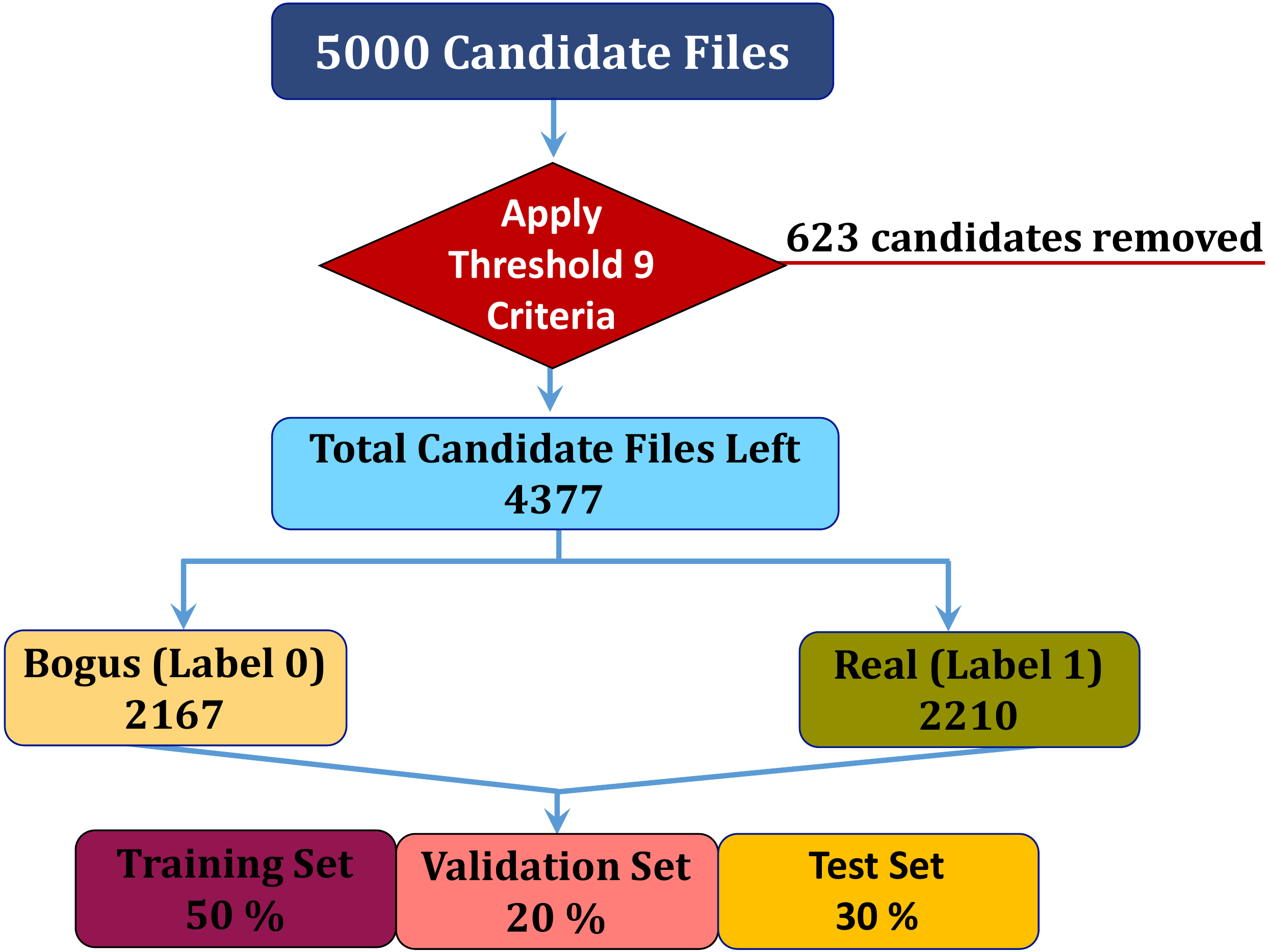}
    \caption{An example of the procedure for selecting the candidates for training the CNN using the thresholding approach, applied on the 5000 candidates. In this example, \textit{Atleast 9 (T9)} is applied and we note that 623 candidates are discarded. Then, the remaining candidates (4377) are split into training, validation and test set for training and evaluation processes.
    \label{fig:thresholding_process}}
\end{figure}

\subsection{\textbf{Labelling data with Thresholding Method}}\label{subsec:data_labelling}

Our sample of real-bogus training samples is constructed from a pool of 5000 candidates that has been selected randomly. Each object is classified by 10 vetters as either real, bogus or can be skipped if they are unsure how to classify it. Each vetter's ability to classify a particular candidate may vary according to the class, images, criteria and experience. This may result in very different classifications for the same candidate. We therefore assign a probability $\mathcal{P}\left(Real\right)$ and $\mathcal{P}\left(Bogus\right)$ to each vetted candidate as follows:

\begin{equation}
    \mathcal{P}\left(\textrm{Real}\right)=\frac{n\left(R\right)}{n\left(T\right)},
	\label{eq:probability_real}
\end{equation}

\begin{equation}
    \mathcal{P}\left(\textrm{Bogus}\right)=\frac{n\left(B\right)}{n\left(T\right)},
	\label{eq:probability_bogus}
\end{equation}

\noindent where $n\left(R\right)$ is the total number of vetters who classified a candidate as real, $n\left(B\right)$ is the total number of vetters who classified a candidate as bogus, $n\left(T\right)$ is the total number of vetters classifying a particular candidate and in this case $n\left(T\right)=10$ as none of the vetters skipped a particular candidate. The vetters classification results are illustrated in a bar plot as shown in Figure \ref{fig:Consistency_in_labelling_5000}. If a candidate has $\mathcal{P}\left(Real\right) \geq 0.9$, it will be given the label real or if $\mathcal{P}\left(Bogus\right) \geq 0.9$, it will be assigned as bogus in the bar plots with x-axis ``$\textit{Atleast}$ 9'' ($\textbf{T9}$). Each bar corresponds to a threshold, e.g, the last bar indicates that out of 5000 candidates, 3216 labelled objects are agreed upon by all the 10 vetters, of which 1572 are bogus and 1644 real. Therefore, there are 1784 sources remaining for which vetters did not agree completely and these confused candidates are removed from the data when using \textbf{T10}. In the ``$\textit{Atleast 9}$'' (\textbf{T9}) case, there are 2167 sources where vetters agreed they are bogus and 2210 sources where ``$\textit{Atleast 9}$'' vetters say they are real, and so on. Going down to $\textbf{T5}$, that is, ``$\textit{Atleast 5}$''  vetters where all the 5000 candidates have been assigned a class, with 2384 bogus and 2616 real. In this paper, we will analyse what happens to the classification results when varying thresholds from \textbf{T8} to \textbf{T10}.

\subsection{\textbf{Labelling data with Latent Class Model, $\textrm{L}_{lcm}$}}\label{subsec:LCM}
Latent class model (LCM) is a statistical technique used to classify candidates into mutually exclusive, or latent classes. When data is in the form of a series of categorical responses, for example individual-level voting data as in the case of real-bogus classification, it is often an interesting technique to identify and characterize clusters of similar cases. In this paper, some confused sources\footnote{5 vetters labelled them as bogus and the other 5 as real.} were removed from the data when using the \textit{thresholding} method as shown in Figure \ref{fig:thresholding_process}. The process outlined in Figure \ref{fig:thresholding_process}, is exactly the same irrespective of the threshold used, the only difference lies in the number of candidates removed when the threshold is applied. However, these confused sources are useful for determining how the system will perform in a real-world scenario. Therefore, confused examples will also  be used during the evaluation phase, and this is achieved by using $\textrm{L}_{lcm}$ to assign them their most likely labels. Therefore, for the $\textrm{L}_{lcm}$ technique we used all 5000 candidates during the training and evaluation phase.

LCM relates a set of observed multivariate variables to a set of latent variables. The latent variable is usually discrete. A class is identified by a pattern of conditional probabilities that provide the chance that variables are given certain values.

Let us take the situation of real versus bogus, we want to use LCM to understand the labels provided by the vetters and give a final label to each source. Imagine that class 0-1 is given to a range of candidates with characteristics a, b, c, and d and that class 0 is associated with the presence of characteristics a, b, and c, and class 1 with characteristics b, c and d. LCM will try to detect the presence of latent classes (the candidates entities), generating patterns of association in the characteristics. Then LCM is used to classify candidates according to their maximum likelihood class membership.

The introduction of a latent variable ensures conditional independence within each latent class, the observed variables, in this case the vetters' labelling, are statistically independent. The association between the observed variables is explained by the classes of the latent variable \citep{mccutcheon1987latent}.

The latent class model can be formulated as follows:

\begin{equation}
    \mathcal{P}_{i_{1}, i_{2}, \dots, i_{N}} \approx \sum_{c}^{C}\mathcal{P}_{c}\prod_{n}^{N}\mathcal{P}_{i_{n},t}^{n},
	\label{eq:LCM}
\end{equation}

\noindent where $C$ is the number of latent classes and in our case, $C = 2$, i.e. real and bogus classes. $N$ is the number of observed binary variables (in this case, $n = 1, \dots, 10$, since we have 10 vetters) and $\mathcal{P}_{c}$ are the unconditional probabilities that should sum to one. $\mathcal{P}_{i_{n},t}^{n}$ are the marginal/conditional probabilities.

\begin{figure*}
\centering
\subfloat[\texttt{MeerCRAB1}]{\includegraphics[width=0.8\textwidth]{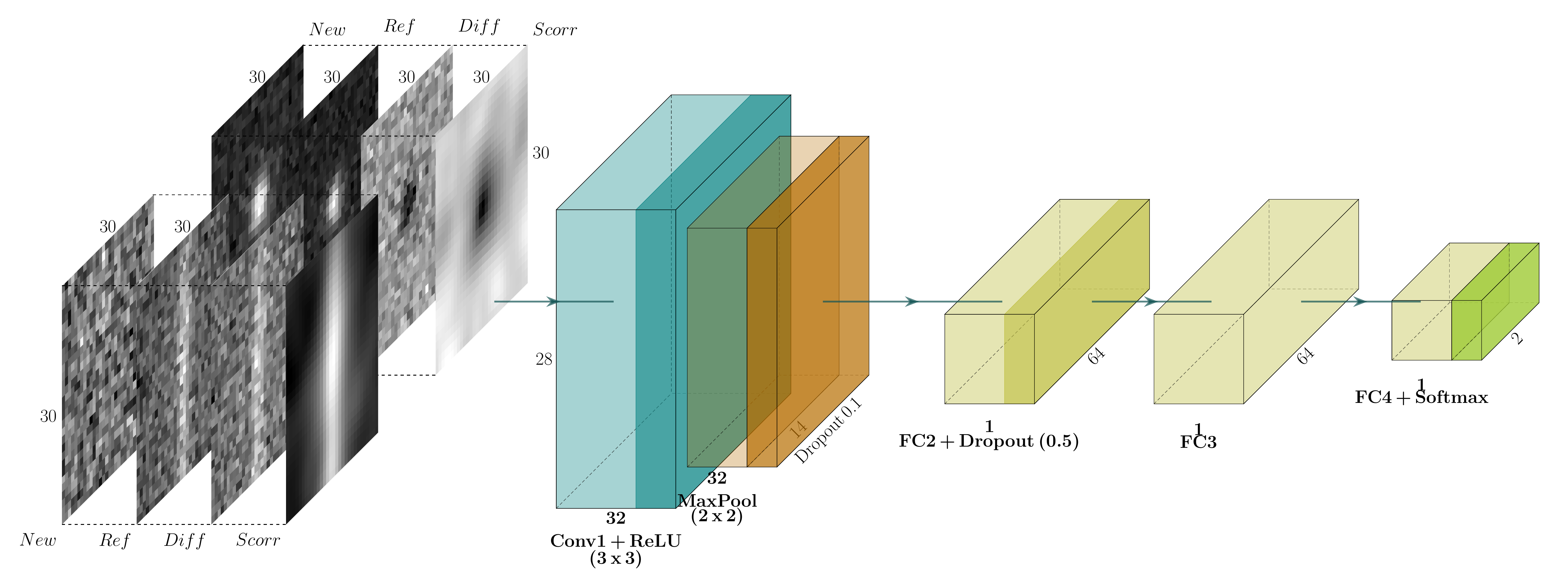}}\\
\subfloat[\texttt{MeerCRAB2}]{\includegraphics[width=0.9\textwidth]{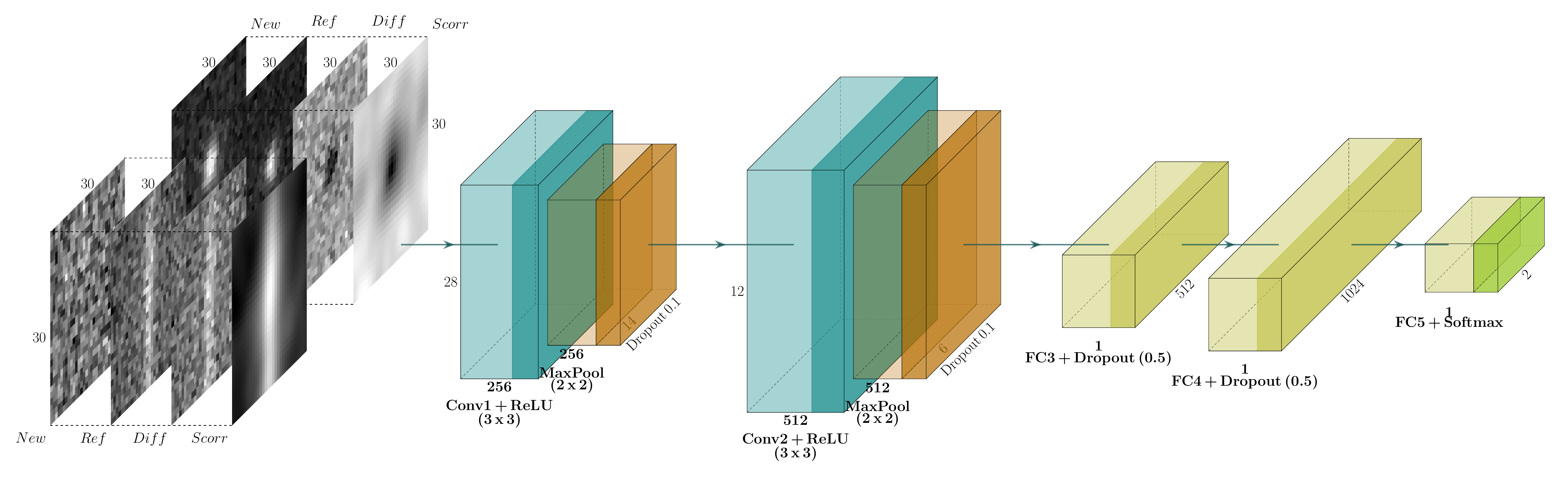}}\\
\subfloat[\texttt{MeerCRAB3}]{\includegraphics[width=\textwidth]{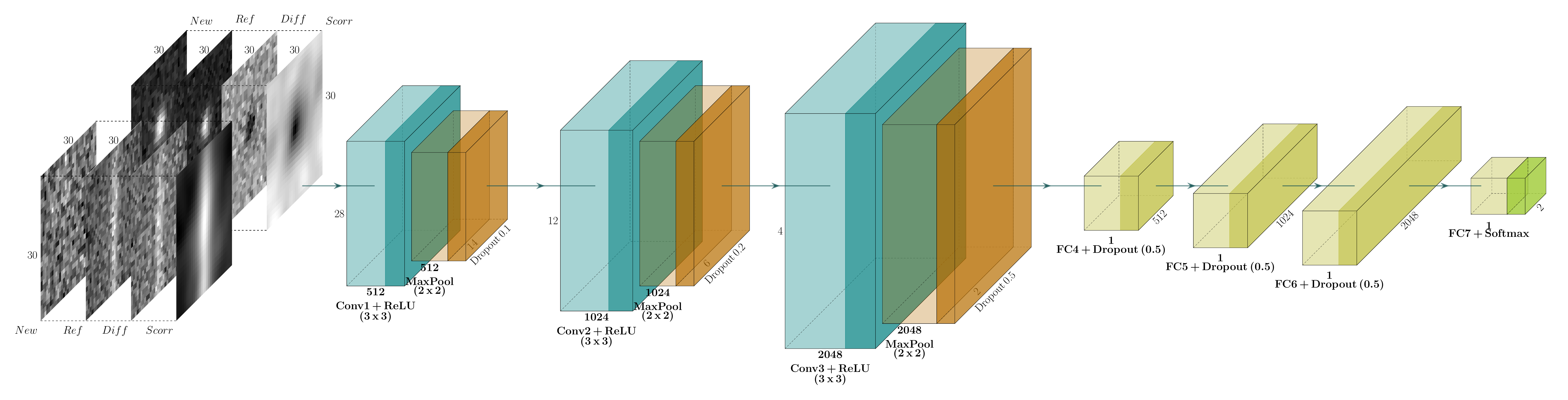}}
\caption{\label{fig:network-architecture} The three network architectures considered in this work: \texttt{MeerCRAB1}, \texttt{MeerCRAB2} and \texttt{MeerCRAB3}. We show four images grouped together (new, reference, difference and significance) to form the input of the networks, followed by convolutional layers, max-pooling, dropout and dense layers. At the end, the network outputs a probability whether a candidate is either real or bogus during the prediction phase.}
\end{figure*}

\begin{figure*}
\centering
\subfloat[Accuracy]{\includegraphics[width=0.50\textwidth]{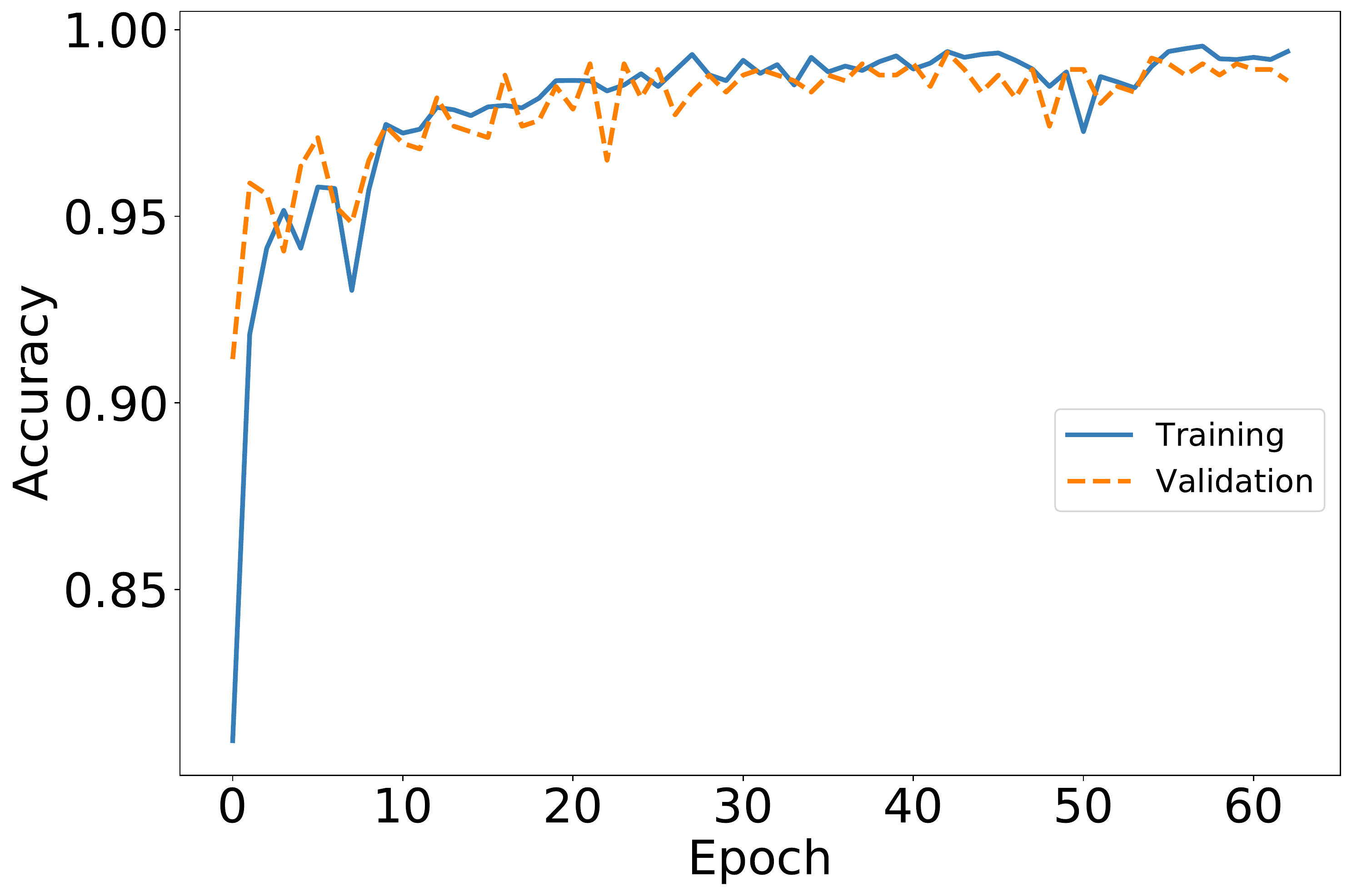}}
\subfloat[Loss]{\includegraphics[width=0.50\textwidth]{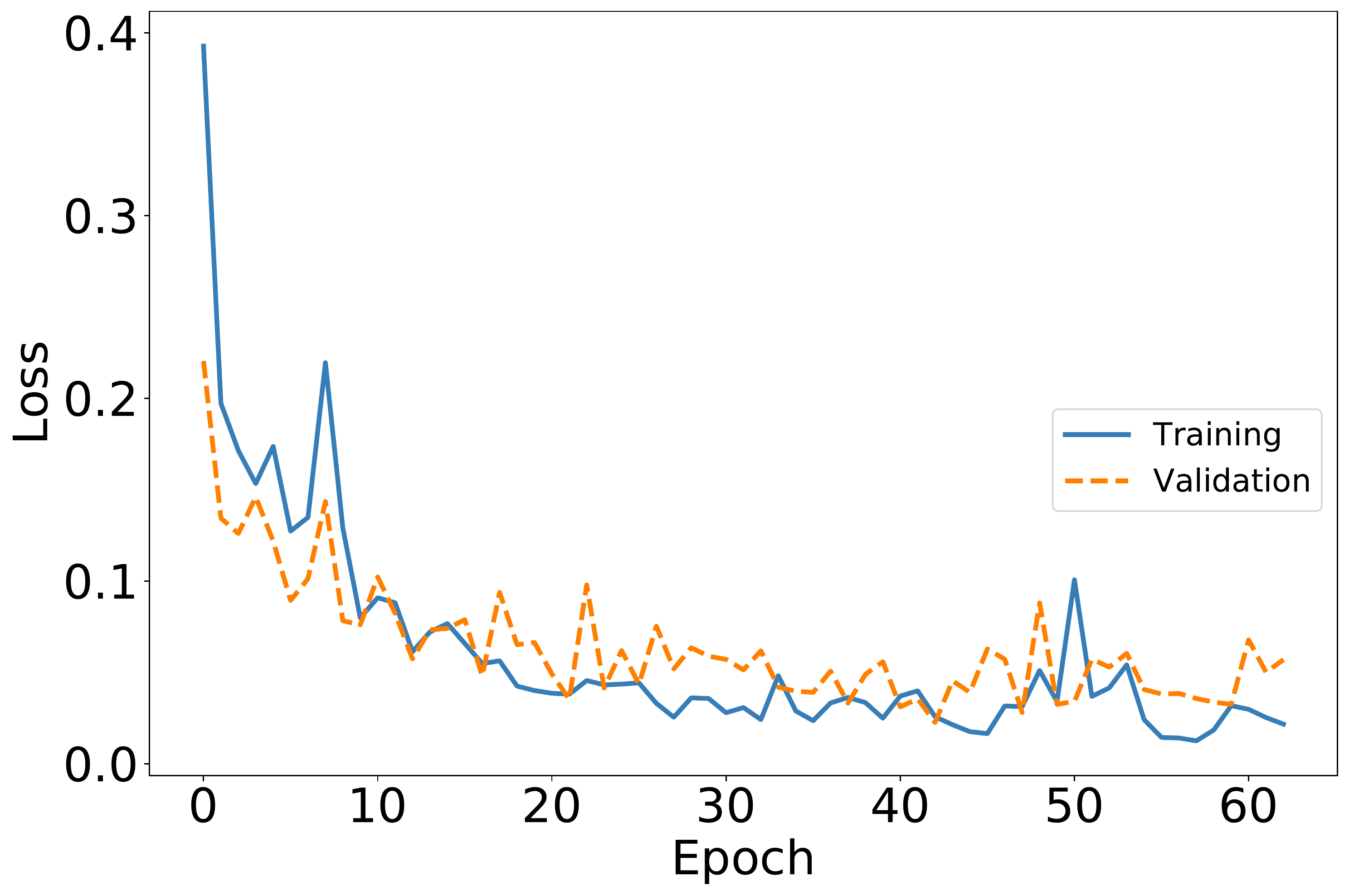}}
\caption{\label{fig:optimization_curves}Learning curves of the \texttt{MeerCRAB3} model with \textit{T9} and $NRD$ as input. The left panel shows the training and validation accuracy over iterations/epochs. The right panel shows the change in negative log-likelihood/loss with epochs. It can be observed that the training objective decreases consistently over iterations, but at some point (around 48 epochs) the validation set loss eventually starts to increase again. An early-stopping technique is applied to avoid overfitting by terminating that training process. At this stage, the algorithm picks the best parameters at 48 epochs.}
\end{figure*}

\section{\texttt{MeerCRAB}: A Real-Bogus Intelligent Distinguisher for the MeerLICHT facility using Deep Learning}\label{sec:MeerCRAB-pipeline}

%\ZH{Machine Learning (ML) can be broadly categorized as \textit{supervised} or \textit{unsupervised}. ML has led to a paradigm shift in astronomy \ZH{and other fields} for more than two decades. Its application can be observed in various studies, namely, pulsar sifting using  ML \citep{2020arXiv200714843L}, a deep-learning classifier for Fast Radio Bursts (\texttt{FETCH}, \citealt{10.1093/mnras/staa1856}) and point source detection (\texttt{DEEPSOURCE}, \citealt{2019MNRAS.484.2793V}) to name a few. Concomitantly, Deep Learning (DL, \citealt{PMID:17921042}) and Convolutional Neural Networks (CNNs, \citealt{Lecun99objectrecognition}) have solved many complex image and speech recognition problems thereby revolutionising how ML is practically applied (\citealt{2014arXiv1406.2661G,2019MNRAS.485.5831L}).}

Before providing further details on the model used, we define a few important terms that will be used in this paper. We labelled bogus examples as 0 and real examples as 1, therefore,

\begin{itemize}
	\item{TP: true positives are real candidates correctly classified as real,}
	\item{TN: true negatives are bogus candidates correctly classified as bogus,}
	\item{FP: false positives are bogus candidates that are classified as real,}
	\item{FN: false negatives are real candidates that are classified as bogus.}

\end{itemize}

From the above definition, our classifier must minimize the number of FP and FN. This is because we do not want to lose many real candidates falsely classified as bogus (FN) and we want to minimize contamination (FP) by any bogus candidates in our final sample.

To overcome this challenge, we employed a Convolutional Neural Network as it has been proven by various studies to have excellent classification performance \citep{2017MNRAS.472.3101G, 2017ApJ...836...97C, 2019PASP..131a8002B, 2019MNRAS.484.2793V,2020arXiv200714843L}. In this work, we construct three CNN models, referred to as \texttt{MeerCRAB1}, \texttt{MeerCRAB2}, and \texttt{MeerCRAB3}. The details of the CNN models are illustrated in Figure \ref{fig:network-architecture}. \texttt{MeerCRAB1} is a network with a single convolutional layer (\textit{CL}), \texttt{MeerCRAB2} comprises two \textit{CLs} while \texttt{MeerCRAB3} is a deeper network with three \textit{CLs}. Each \textit{CL} is made of $\left( 3 \times 3 \right)$ pixel filters, together with a Rectified Linear Unit (ReLU, \citealt{2018arXiv180308375A}) function, followed by a pooling layer with filter size of $\left(2\times2\right)$. After the \textit{CLs}, we used fully connected layers (also known as dense layers). In addition, we use a dropout rate varying from (0.1 to 0.5) after each of the pooling and dense layers as seen in Figure \ref{fig:network-architecture} to avoid over-fitting. For the output layer, we used a softmax function that outputs a probability value between 0 and 1.

The implementation of \texttt{MeerCRAB} is made using the \texttt{Tensorflow}\footnote{\href{https://github.com/tensorflow/tensorflow}{https://github.com/tensorflow/tensorflow}} and \texttt{Keras} \citep{2018ascl.soft06022C} API with \textit{Python v3.6}. For training the \texttt{MeerCRAB} models, we used an Nvidia GeForce GTX 1080Ti 11GB GPU. During training, the binary cross-entropy loss function, Adam optimizer \citep{2014arXiv1412.6980K} with a low learning rate (lr = 0.0002) and a batch-size of 64 were used. We then split our data for each and every experiment as follows: 50\% training, 20\% validation and 30\% testing. As input to the \texttt{MeerCRAB} models, we cropped the $\left(100\times100\right)$ pixel images that were analysed by vetters to $\left(30\times30\right)$ pixel images. We also utilized the $\left(100\times100\right)$ pixel images during training, but we observed a drop in performance and therefore only use the $\left(30\times30\right)$ pixel images cropped from the centre. This pre-processing step increases the likelihood that the models will retain useful information and not be distracted by noise or spurious patterns.

As described above, the MeerLICHT database provides four images $N$, $R$, $D$ and $S$. We perform several analyses which are presented in \S\ref{sec:results_and_analysis}, where we use a combination of these images. This work allows us to understand which image/s is/are important for helping the CNN to output better classification results. We therefore use cut-outs of $N$, $R$, $D$, $S$ and stack them to form either of these input images: $NRDS$ $\left(30\times30\times4\right)$, $NRD$ $\left(30\times30\times3\right)$, $NR$ $\left(30\times30\times2\right)$, $D$ $\left(30\times30\times1\right)$, or $S$ $\left(30\times30\times1\right)$.

Afterwards, we apply data augmentation to mitigate the risk of over-fitting which may result in poor generalisation performance upon data outside the training set. Therefore, at each training step, the images are augmented by flipping them randomly in a horizontal and/or vertical direction. We do not apply any rotation or translation to the images. This data augmentation step is important as it helps to increase the training sample size and the probability that the CNN models will encounter similar images twice, will decrease.

Moreover, to further avoid any over-fitting during training, we employ an early stopping technique to stop the training process if no further decrease in validation loss is observed for several epochs. The various cases and models are trained for a number of epochs varying from 40 to 150. We show the optimisation curves (the accuracy and loss curves) during training and validation in Figure \ref{fig:optimization_curves} for the best models with $NRD$ and \textbf{T9} as input. We observe that both the training and validation accuracy reaches a range between 98.5\% to 99.5\% where the algorithm picks the best parameters at 48 epochs.

\subsection{\textbf{Evaluation Performance}}\label{subsec:evaluation_metrics}
We use different evaluation techniques, for instance, the accuracy, the precision, the recall and the Matthew correlation coefficient (MCC) metrics to evaluate the \texttt{MeerCRAB} models. In addition, we utilize the McNemar test for model comparison.

\begin{table}
	\centering
	\caption{Contigency table used as a visual aid for model selection. The sum of A, B, C, and D represents the total number of instances in the test set. A is the number of instances correctly classified by both models. D is the number of instances misclassified by both models. B is the number of objects that model 1 correctly classified, but has been misclassified by model 2. C is the number of examples misclassified by model 1 but being correctly classified by model 2.}
	\label{tab:Contigency_table_explain}
	\begin{tabular}{ccc} % four columns, alignment for each
		\hline
		                & Model 2 Correct & Model 2 Wrong\\
		\hline
		Model 1 Correct & A & B\\
		Model 1 Wrong   & C & D\\
		\hline
	\end{tabular}
\end{table}

\begin{figure}
	\centering
	\includegraphics[width=0.5\textwidth]{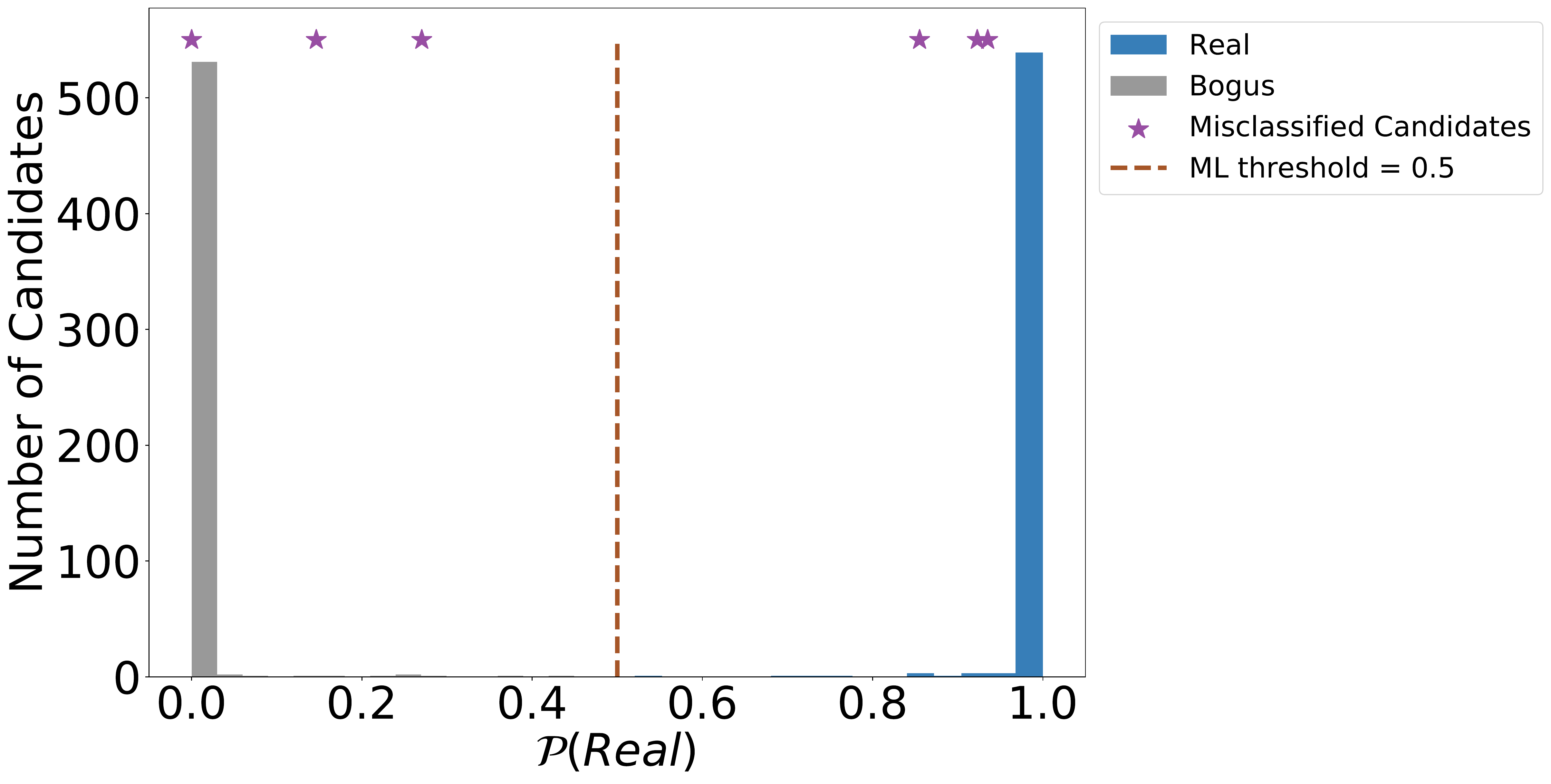}
    \caption{Probability distribution on test data with 1095 candidates, trained with model configuration \texttt{MeerCRAB3} with \textit{T9} and $NRD$ as input.}
    \label{fig:probability_distribution}
\end{figure}

\subsubsection{\textbf{McNemar's Test}}\label{subsub:McNemar_test}
The McNemar test \citep{mcnemar1947note} is a statistical test used to check marginal homogeneity in the context of statistical models. It is used to compare the predictive accuracy of two models' predictions and it is based on a contigency table as shown in Table \ref{tab:Contigency_table_explain}. The latter provides insights for model selection, in contrast to a typical confusion matrix. It shows the number of instances/predictions model 1 and model 2 got right or wrong given a fixed test set. In McNemar's test, a null hypothesis, $H_{0}$ is formulated such that $\mathcal{P}\left(B\right)$ and $\mathcal{P}\left(C\right)$
are similar or it can be interpreted as two models perform equally well. Therefore, the alternative hypothesis $H_{1}$ is that $\mathcal{P}\left(B\right) \neq \mathcal{P}\left(C\right)$ or the two models do not perform equally well. \citet{edwards1948note} proposed a corrected McNemar test statistic that can be computed as given below: 

\begin{equation}
\chi^{2} = \frac{\left(|B-C|-1\right)^{2}}{B+C}.
\end{equation}

\noindent If the sum of B and C is greater than 25 or sufficiently large, under $H_{0}$ the $\chi^{2}$ value follows a chi-squared distribution with one degree of freedom. If we set a significance threshold for example, $\alpha=0.05$, the p-value can be computed. Assuming that the null-hypothesis is true, the p-value implies the probability of observing a larger chi-squared value. However, if the p-value is less than $\alpha$, then the null hypothesis is rejected, that is, the two models do not perform equally well.

In the case where the sum of B and C is less than 25, an exact binomial test is used instead, since the chi-squared value may not be well approximated by the chi-squared distribution. Thus, the exact p-value is calculated as follows:

\begin{equation}
\textrm{p-value} = 2 \times \sum_{i=B}^{n} \binom{n}{i} 0.5^{i}(1-0.5)^{n-i},
\end{equation}

\noindent where $n=B+C$ and a factor 2 indicates the computation of a two-sided p-value. For model selection, if the p-value is less than 0.05, we reject the null hypothesis, thus one model is outperforming the other. However, if the p-value is greater than 0.05, we do not reject the null hypothesis and it indicates that both models perform equally well.

\section{Results and Analysis}\label{sec:results_and_analysis}

\begin{figure}
	\centering
	\includegraphics[width=0.5\textwidth]{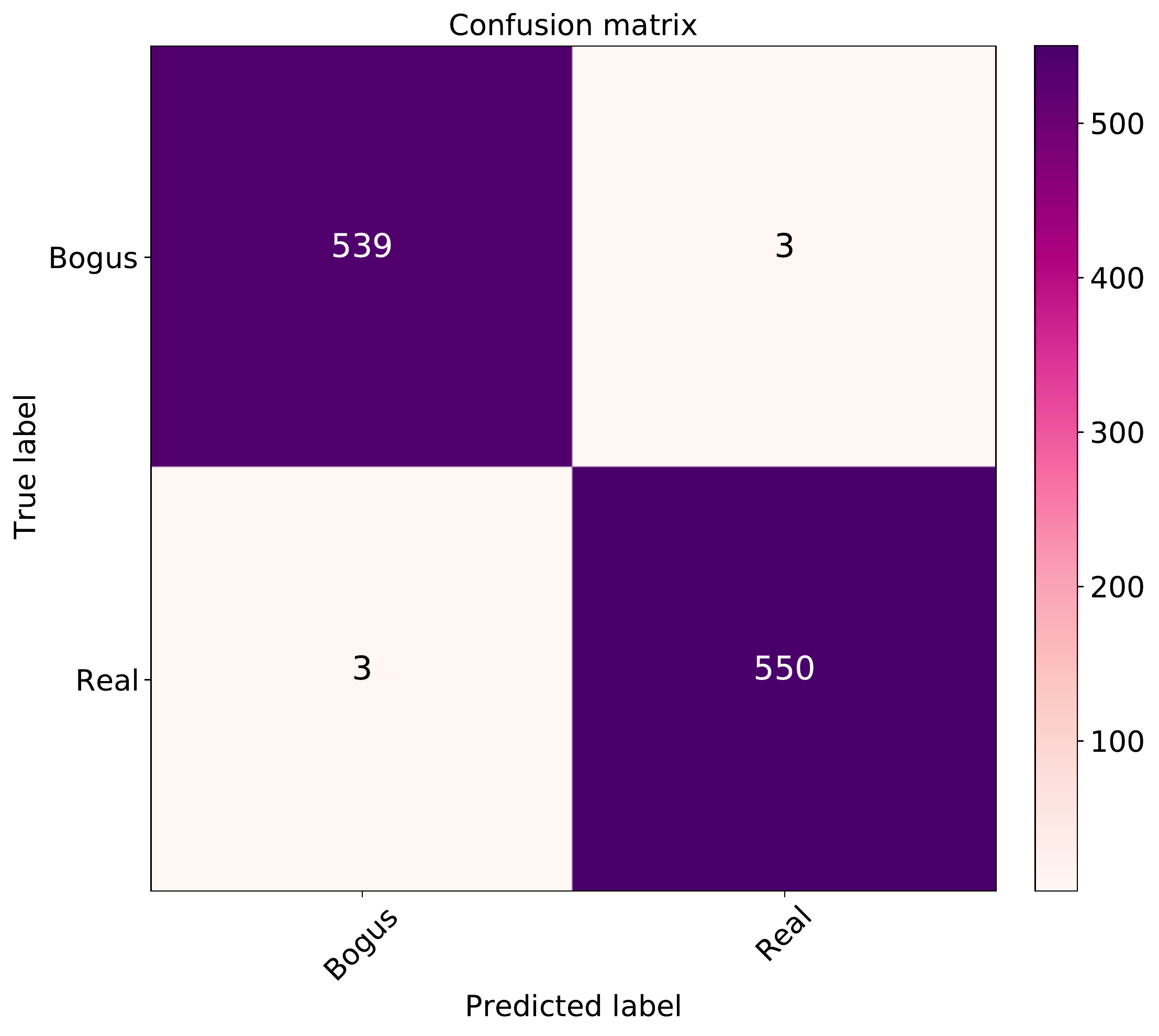}
    \caption{Confusion Matrix without normalisation from a test set of 1095 candidates. The model is trained using \texttt{MeerCRAB3} with \textit{T9} and $NRD$ as input. The diagonal represents the correctly classified instances in the test. The off-diagonals represent the number of instances that are misclassified. We note that we have a very low FP and FN with this model configuration.}
    \label{fig:confusion matrix}
\end{figure}

\begin{figure*}
\centering
\subfloat[Bogus feature maps]{\includegraphics[width=0.45\textwidth]{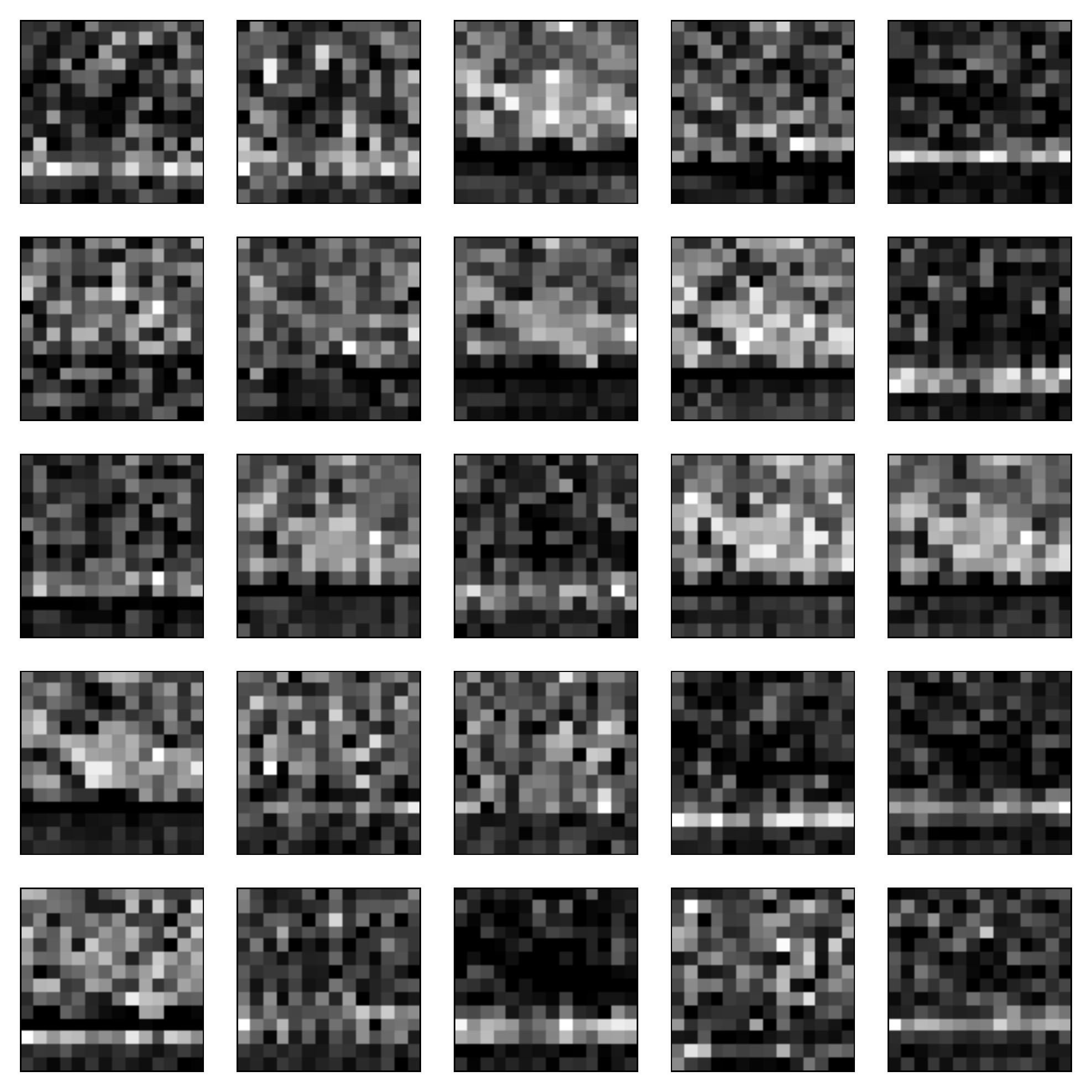}}
\subfloat[Real feature maps]{\includegraphics[width=0.45\textwidth]{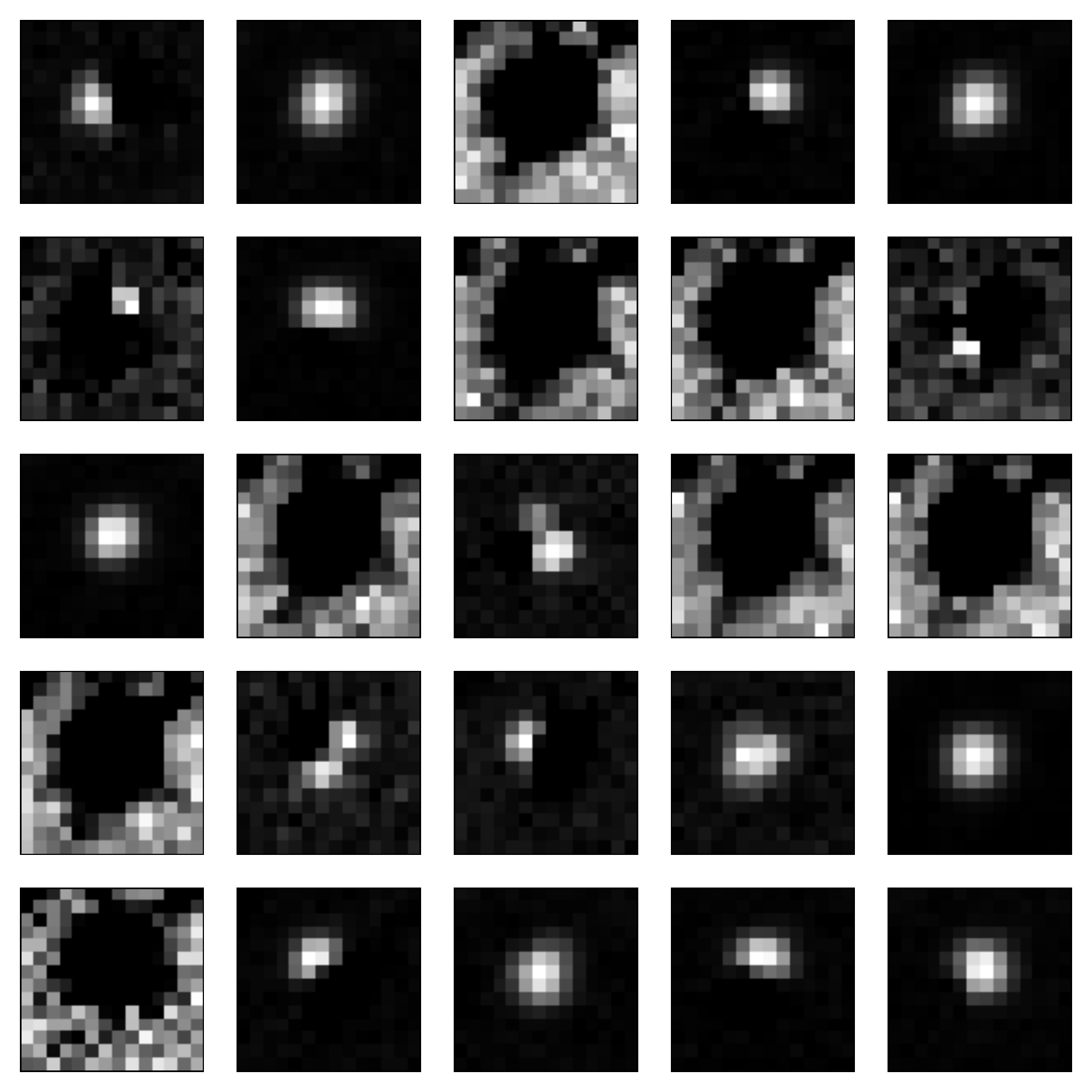}}
\caption{\label{fig:feature_maps}Feature maps induced by the convolutional layer in \texttt{MeerCRAB3} given a ``bogus'' and ``real'' source, respectively. We observe that the model distinguishes between background noise and the source at the center of an image.}
\end{figure*}

The classification performance of the \texttt{MeerCRAB} models are analysed as follows:
\begin{itemize}
	\item{The \texttt{MeerCRAB} models are trained on a subset sample of input images and validated on an unseen image sample.}

	\item{During the prediction phase, the trained models are then used to output probabilistic predictions for unseen images in the test set as shown in Figure \ref{fig:probability_distribution}. The probability distribution from the output of the \texttt{MeerCRAB} models spanned the range of $\mathsf{P}_{\texttt{MeerCRAB}} \in \big[0;1]$. Therefore, a candidate is predicted as bogus if $\mathsf{P}_{\texttt{MeerCRAB}} < 0.5$ and as real if $\mathsf{P}_{\texttt{MeerCRAB}} \geq 0.5$. $\mathsf{P}_{\texttt{MeerCRAB}} = 0.5$ indicates a random guess and the \texttt{MeerCRAB} models are confused between real and bogus candidates.}
\end{itemize}

We investigated various scenarios for training and evaluating the pipeline. We made use of three network structures: \texttt{MeerCRAB1}, \texttt{MeerCRAB2}, \& \texttt{MeerCRAB3}. We also varied the number of images used as input to the three model architectures. We use these combinations of input images $NRDS$, $NRD$, $NRS$, $NR$, $D$ and $S$ independently. In addition, we investigated the effect of varying the thresholding applied on data labelling, ie., the effect of noisy data labels as discussed in \S\ref{subsec:data_labelling} and \S\ref{subsec:LCM}.

For all the experiments considered, it is worth mentioning that we train the networks with and without data augmentation. We found that employing data augmentation decreases the number of misclassifications. Therefore, we report results with the models trained with data augmentation only in this paper. Using data augmentation, we found the objective values steadily decreased during the fitting process on both the training set and validation set. Figure \ref{fig:optimization_curves} provides an illustration of the objective functions and we observe that both the training and validation loss are close to each other, indicating that the models did not overfit on the training sample. Hence, the dropout layers as well as various regularization techniques used in the models are effective measures to prevent overfitting.

\subsection{\textbf{Case studies}}
We start by providing an overall comparison of various scenarios used and will subsequently analyse some of the models along with certain modifications in more details. The \texttt{MeerCRAB} models are evaluated on various metrics. Based on the $\mathsf{P}_{\texttt{MeerCRAB}}$ value, we construct a confusion matrix to have an overview of the classification results. The confusion matrices display the fraction of correctly classified candidates as TP, TN along the diagonal. The off-diagonal values in the confusion matrices show the misclassified examples (FP and FN). We also evaluated the models based on precision, recall, accuracy, and MCC as discussed in \S\ref{subsec:evaluation_metrics}. The results for the various scenario cases are summarised in Table \ref{tab:results}, \ref{tab:threshold_8}, \ref{tab:threshold_9}, \ref{tab:threshold_10} and Figure \ref{fig:confusion matrix}. 

\begin{figure*}
\centering
\subfloat[\texttt{MeerCRAB1} vs \texttt{MeerCRAB2}]{\includegraphics[width=0.33\textwidth]{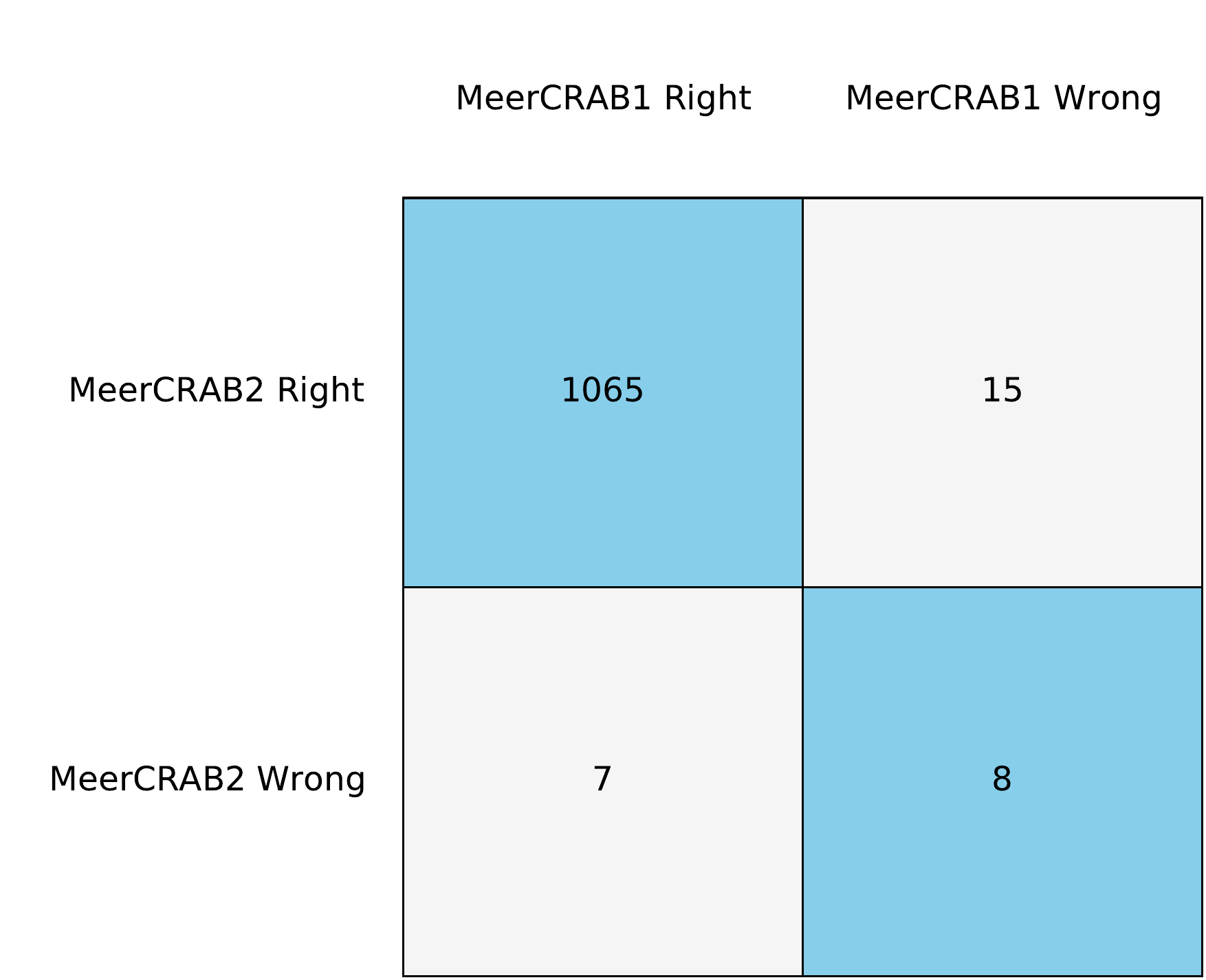}}
\subfloat[\texttt{MeerCRAB2} vs \texttt{MeerCRAB3}]{\includegraphics[width=0.33\textwidth]{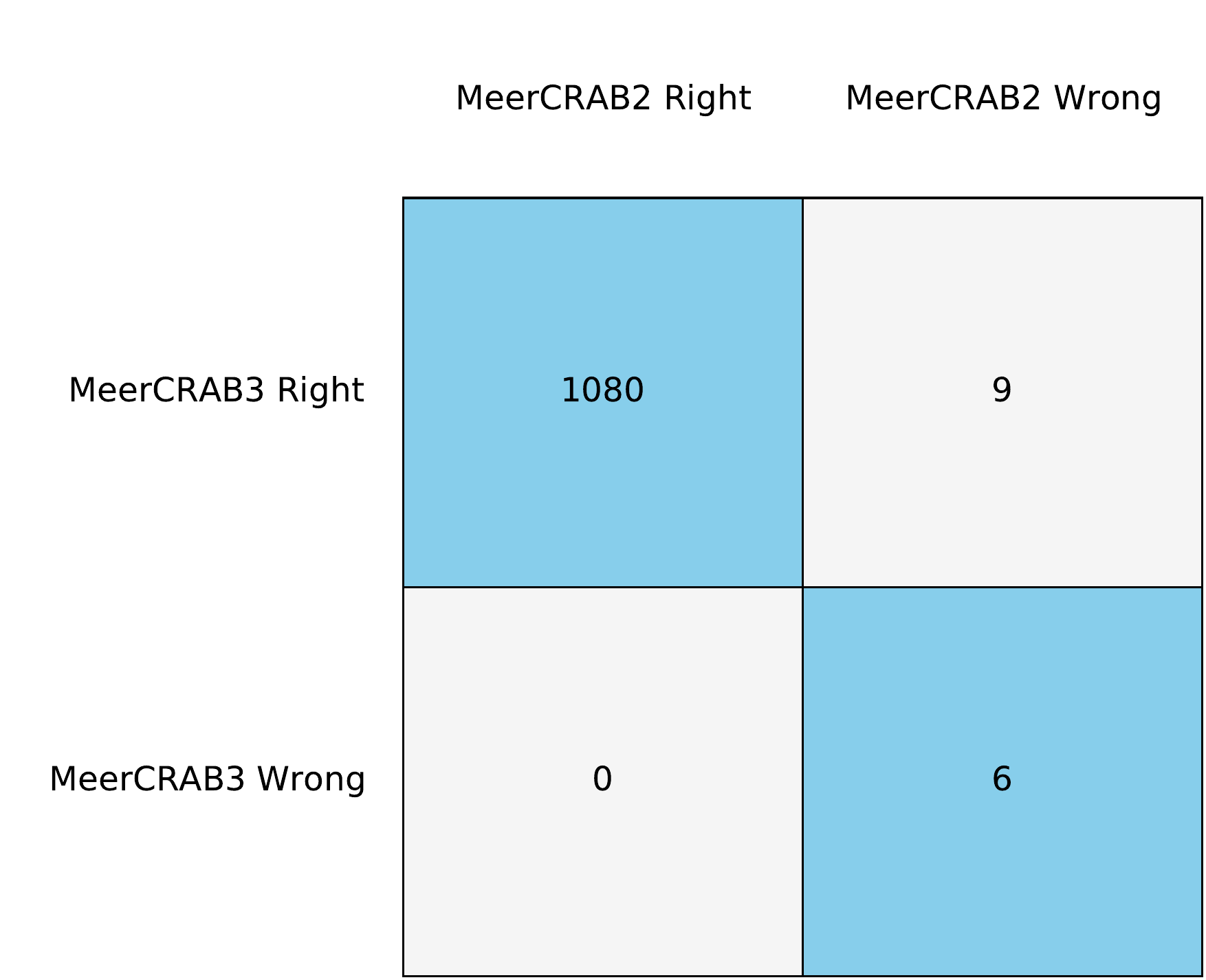}}
\subfloat[\texttt{MeerCRAB1} vs \texttt{MeerCRAB3}]{\includegraphics[width=0.33\textwidth]{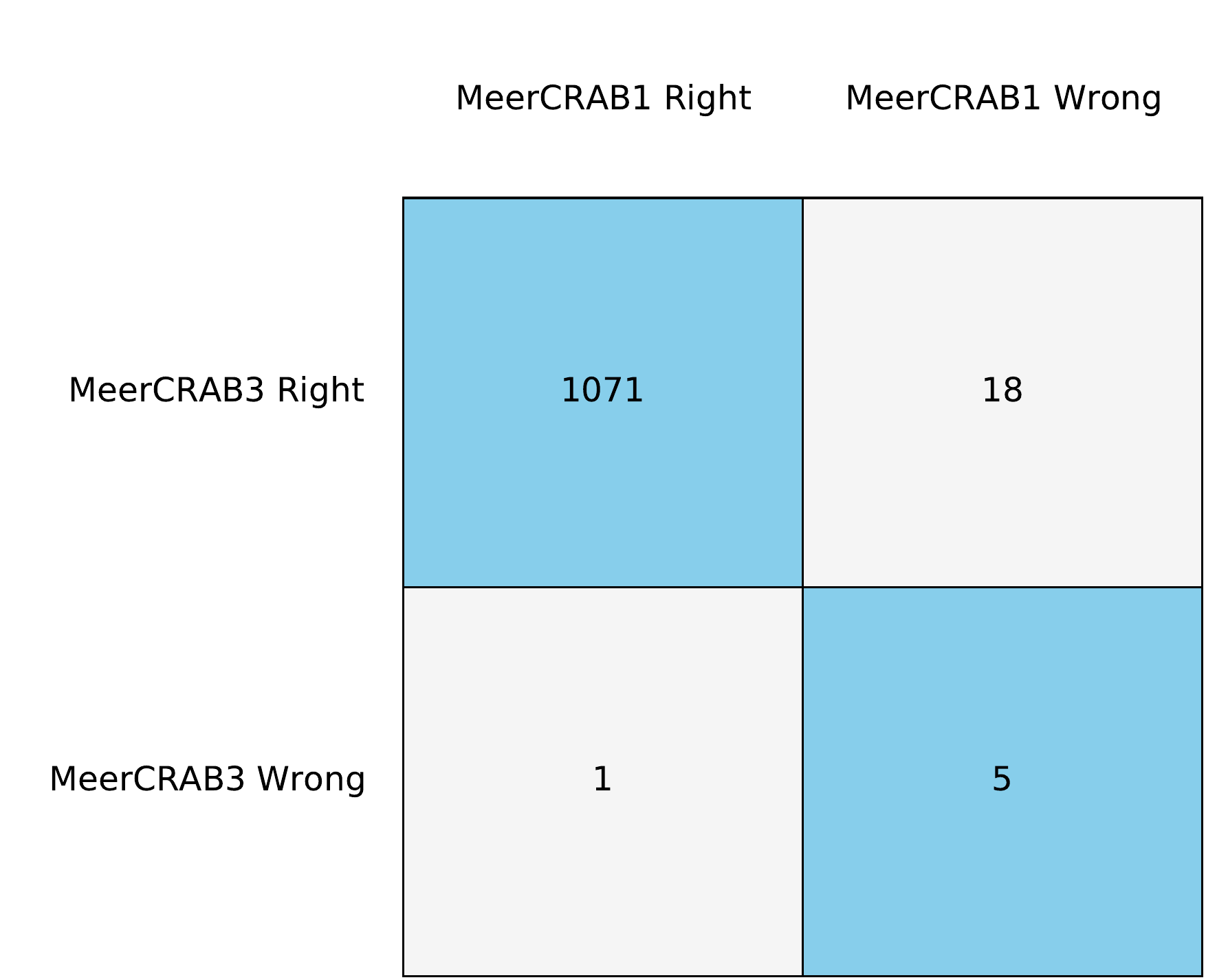}}
\caption{\label{fig:contigency_table}The contigency tables for the three models under consideration with \textit{T9} and $NRD$ as input. For model selection, we compare the performance accuracy of each model on a similar test set. From the plots, we observe that \texttt{MeerCRAB3} is a better model compared to \texttt{MeerCRAB1} and \texttt{MeerCRAB2}.}
\end{figure*}

\subsubsection{\textbf{Data labelling based on Thresholding criteria}}
In supervised learning algorithms, the success of deep neural networks depends highly on the availability and accessibility of high-quality labelled training data. In this work, we found that the presence of label errors (label noise) in the training data greatly reduced the accuracy of all \texttt{MeerCRAB} models on test data. Unfortunately, large training datasets almost always contain examples with inaccurate or incorrect labels. It is a challenging task to train deep neural networks (DNNs) robustly with noisy labels \citep{2018arXiv180406872H} as DNNs have a high capacity to fit noisy labels \citep{2016arXiv161103530Z}, and this results in poorer model performance in practice.

%This leads to a paradox: on the one hand, large datasets are necessary to train better deep networks, while on the other hand, deep networks tend to memorize training label noise, resulting in poorer model performance in practice.

In this paper, in order to sanity check the potential generalisation performance of our models, we utilised latent class models, $\textrm{L}_{lcm}$ to label our data -  including those samples for which vetters could not reach agreement. This approach allows us to introduce labelling noise. Whilst this noise reduces the performance of our models causing accuracy to drop to 0.968, it helps us obtain an impression of real-world performance where imperfect labelling and noise cannot be controlled for. Results show that models trained  using such data still perform very well. This indicates that the networks are robust to noise.

Our analysis also involves comparing various thresholding criteria: \textbf{T8}, \textbf{T9}, and \textbf{T10} along with the \texttt{MeerCRAB3} model with $NRD$ as input. The results for labelling techniques are summarised in Table \ref{tab:results}. We observe that as the threshold increased from \textbf{T8} to \textbf{T10}, the accuracy of the model increases from 0.988 to 0.998 and MCC values increase from 0.976 to 0.995. However, when using the $\textrm{L}_{lcm}$ method, we note a  significant drop in $\textrm{L}_{lcm}$ accuracy (0.968) as deep networks tend to memorize training label noise, resulting in poorer model performance. Therefore, it is necessary to obtain fairly high-quality labels for a CNN to work appropriately, thus removing noisy labelling from the model yields better model performance. 

However, from here on we focus on experimentation that uses labels obtained via thresholding criteria : \textbf{T9} only, and not the $\textrm{L}_{lcm}$. we do this as we aim to have the best optimized model with an adequate number of candidates along with a good level of agreement on the labelling, such that we obtain fewer false positives and false negatives.

\subsubsection{\textbf{Network architectures}}

We trained the \texttt{MeerCRAB} pipeline with three different architectures using \textbf{T9} data. \texttt{MeerCRAB1} consists of a single \textit{CL}, \texttt{MeerCRAB2} is trained with 2 \textit{CLs} and \texttt{MeerCRAB3} with 3 \textit{CLs}. From Table \ref{tab:threshold_8}, Table \ref{tab:threshold_9} and Table \ref{tab:threshold_10}, we note that \texttt{MeerCRAB1} which is a shallow network yields a surprisingly good performance for \textbf{T8} to \textbf{T10}. Looking at Table \ref{tab:threshold_9} with \textbf{T9}, we note that \texttt{MeerCRAB1} achieves an accuracy of 0.980 and MCC = 0.960 on the test set. When using deeper networks (\texttt{MeerCRAB2} and \texttt{MeerCRAB3}), we found that we obtain a better performance. The accuracy for 
\texttt{MeerCRAB2} and \texttt{MeerCRAB3} using $NRD$ as input are 0.986 \& 0.995 and MCC values are 0.973 \& 0.989 respectively.

To have a better understanding of why \texttt{MeerCRAB1} yields a good performance, we plot in Figure \ref{fig:feature_maps} the feature maps of the \textit{CL} for a bogus and real example. We observe that there appears to be feature maps that activate on the background (``dark centre''), while other maps activate on different parts of the centre. This suggests that the network can distinguish between the source itself and the background, thus it is able to classify images relatively unhindered by different levels of noise.

To determine which network performs best, we employ the McNemar statistical test as discussed in \S\ref{subsub:McNemar_test}. We plot the contigency tables for the three models in Figure \ref{fig:contigency_table}. The sample size in the B and C cells are relatively small and $(B+C)<25$ to approximate the chi-square value from the chi-square distribution. We therefore compute the p-value in this case from a binomial distribution. Assuming we use $\alpha=0.05$, if the p-value is less than 0.05, we reject the null hypothesis that both models perform equally well on the test set and if the p-value is greater than 0.05, we do not reject the null hypothesis, we then conclude that the two models perform equally well. From Table \ref{tab:McNemar_test}, we note that when comparing \texttt{MeerCRAB1} with \texttt{MeerCRAB2}, the p-values are greater than 0.05, we therefore conclude that the models have an equal performance. However, when comparing \texttt{MeerCRAB1} with \texttt{MeerCRAB3} and \texttt{MeerCRAB2} with \texttt{MeerCRAB3}, the p-vaue is less than 0.05 this indicates that one model is performing better. When we analyse Figure \ref{fig:contigency_table}(b,c), we note that \texttt{MeerCRAB3} has less instances being misclassified compared to \texttt{MeerCRAB1} and \texttt{MeerCRAB2}. Therefore, we conclude \texttt{MeerCRAB3} is a better model compared to \texttt{MeerCRAB1} and \texttt{MeerCRAB2}.

\subsubsection{\textbf{Input Images}}
In this section, we investigate the effect of adding and removing images from the input to the CNN models. Vetters were shown only three images during vetting: $N$, $R$ and $D$ images. However, when training and evaluating the three networks,  we compare the different groups of images, to see whether a competitive performance can be achieved with more or less input data.  Focusing on \textbf{T9} and \texttt{MeerCRAB3}, we note that $NRD$ input yields the best performing model with an accuracy of 0.995 and MCC value of 0.989. Similar results are obtained with \textbf{T8} and \textbf{T10}. The second best model in \textbf{T9} is with input $NRS$ yielding an accuracy of 0.990 and MCC value of 0.980. With only $NR$ as input, we note that \texttt{MeerCRAB3} performs equally well. Therefore, using a reduced image input set ($NR$) yields a competitive performance and indicates that a reduced set of images is sufficient for separating real and bogus. However, we note that using only $D$ or $S$ as input worsens the classification performance and this indicates that using information only from the difference or significance imaging is insufficient.

\subsection{\textbf{Analysis of misclassification}}

With the various investigations, it is worth mentioning that the classification performance is, in general, very similar for the different \texttt{MeerCRAB} models, i.e., neither a particular network structure nor the involved parameters seem to have a significant influence on the final classification performance. The conclusion one can draw at this point is that  a standard CNN model seems to be well-suited for the task at hand and that even relatively simple networks yield a performance that is competitive with state-of-the-art approaches.

\begin{figure}
	\centering
	\includegraphics[width=\columnwidth]{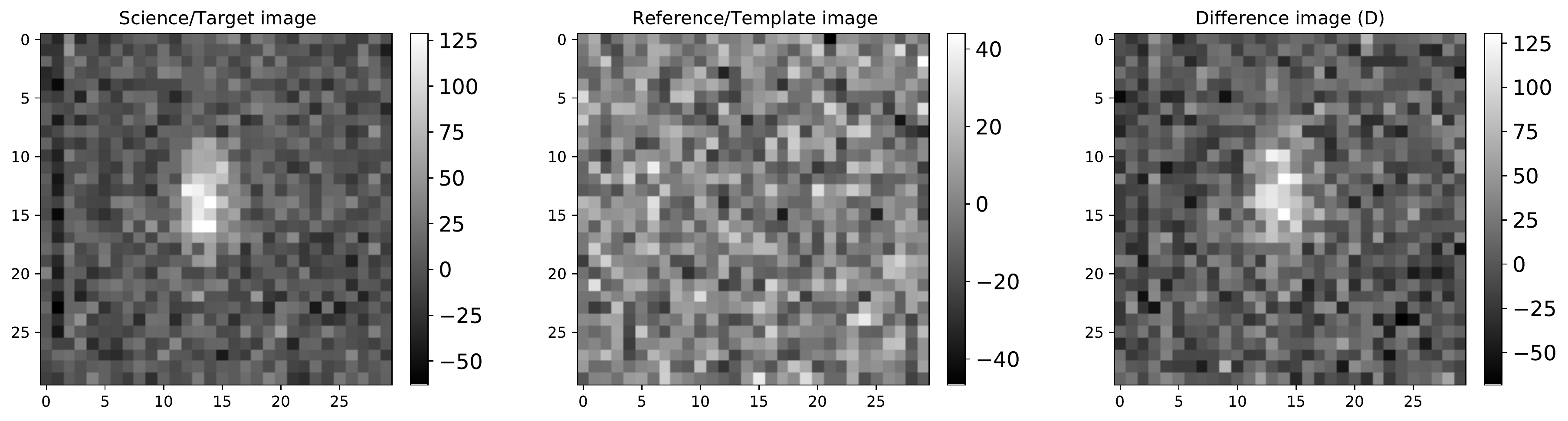}
	\includegraphics[width=\columnwidth]{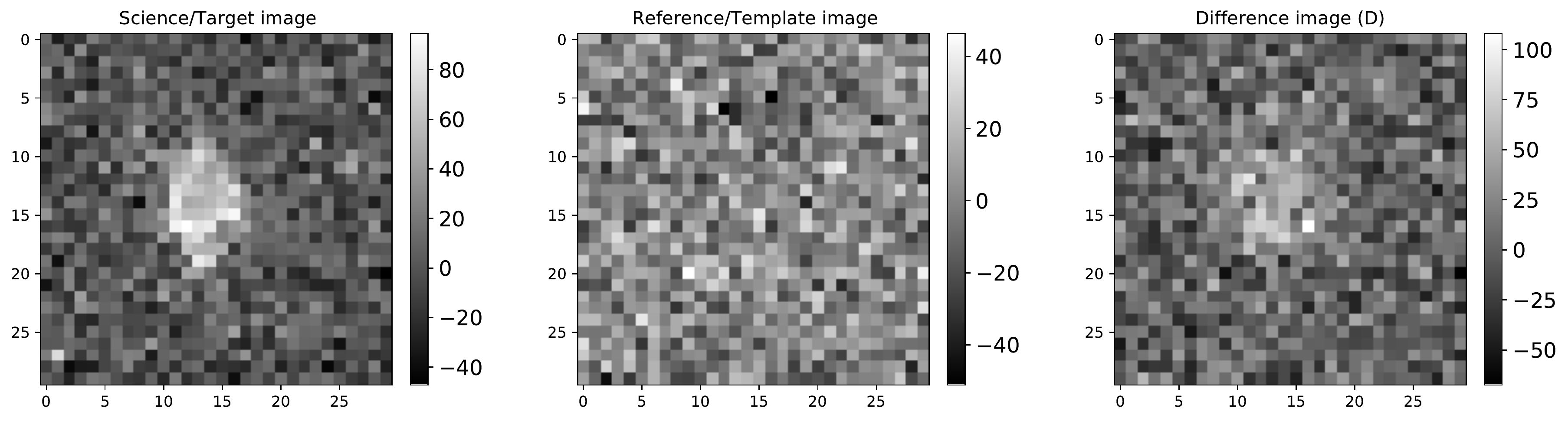}

	\includegraphics[width=\columnwidth]{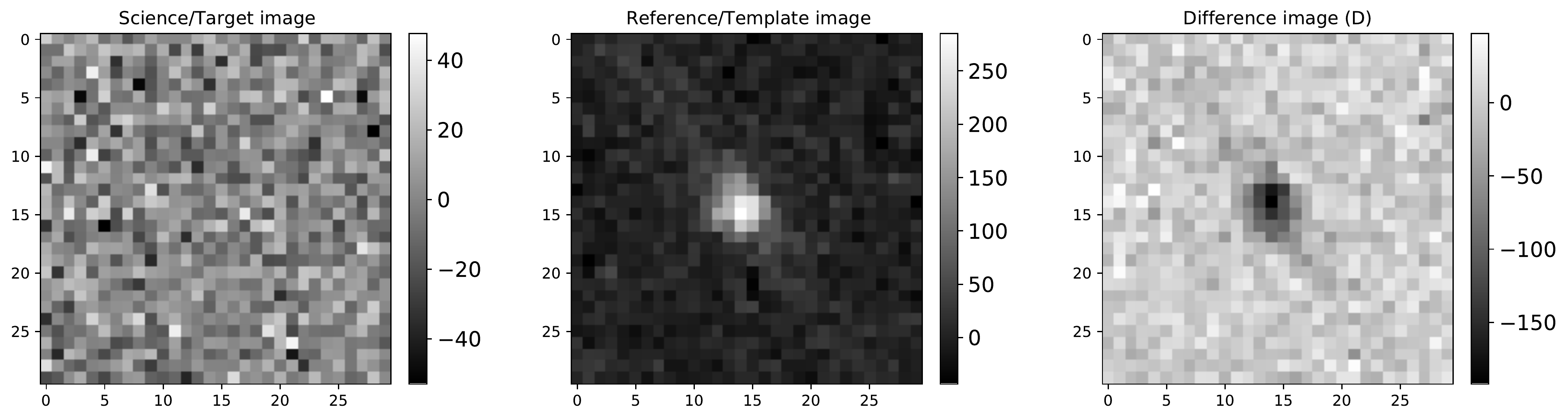}

    \caption{The false positives obtained with \texttt{MeerCRAB3}, trained with \textit{T9} labelling and $NRD$ as input. These candidates are classified as bogus by humans. However, the best performing network misclassified them as real.}
    \label{fig:misclassification_bogus_1}
\end{figure}

\begin{figure}
	\includegraphics[width=\columnwidth]{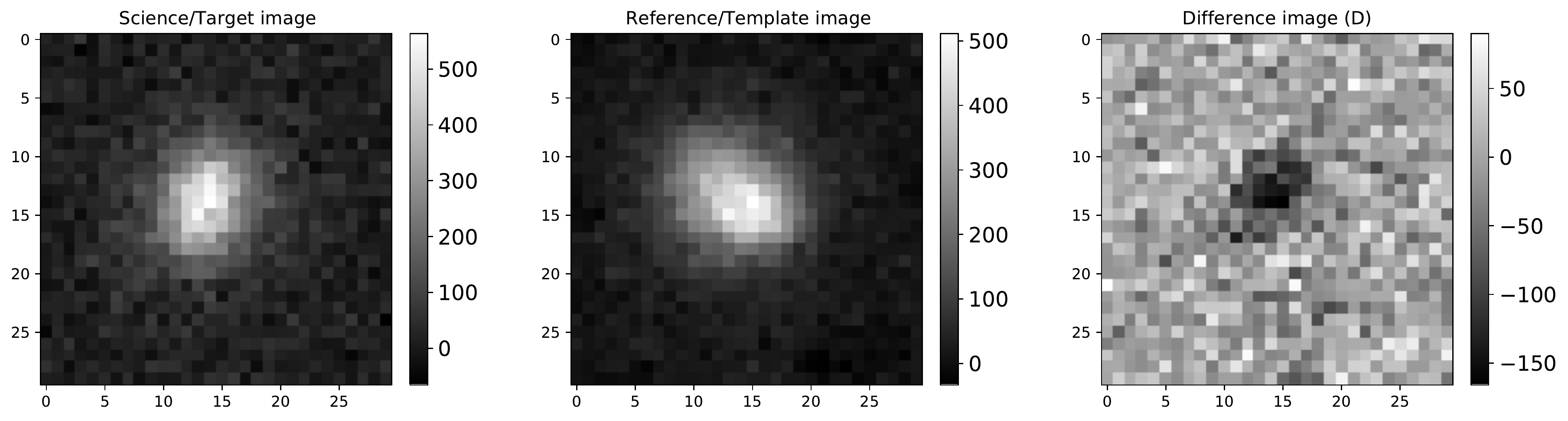}
	\includegraphics[width=\columnwidth]{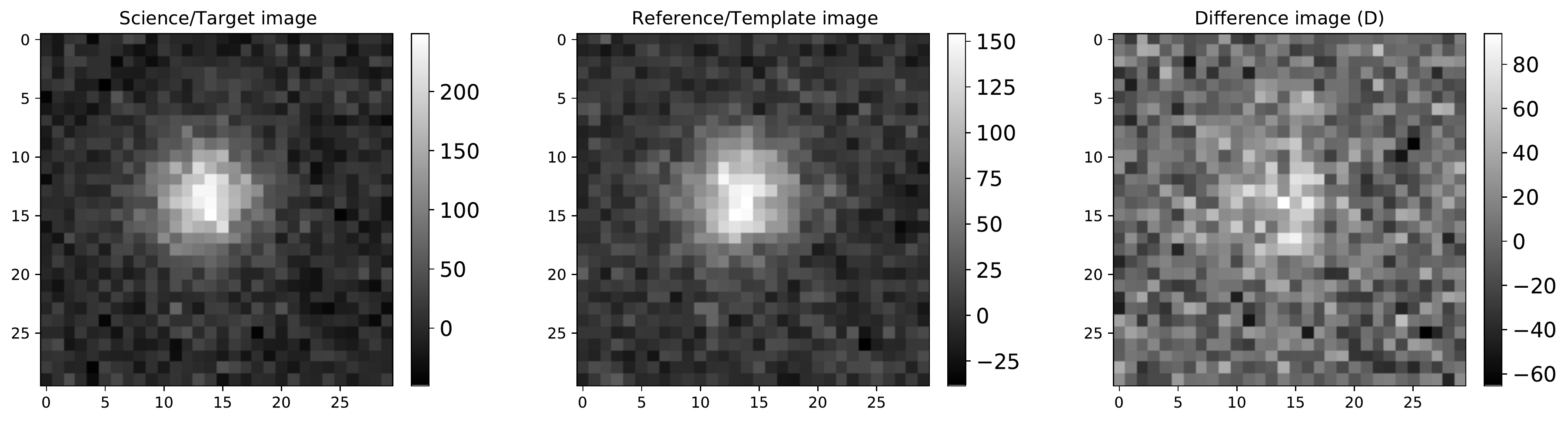}
	\includegraphics[width=\columnwidth]{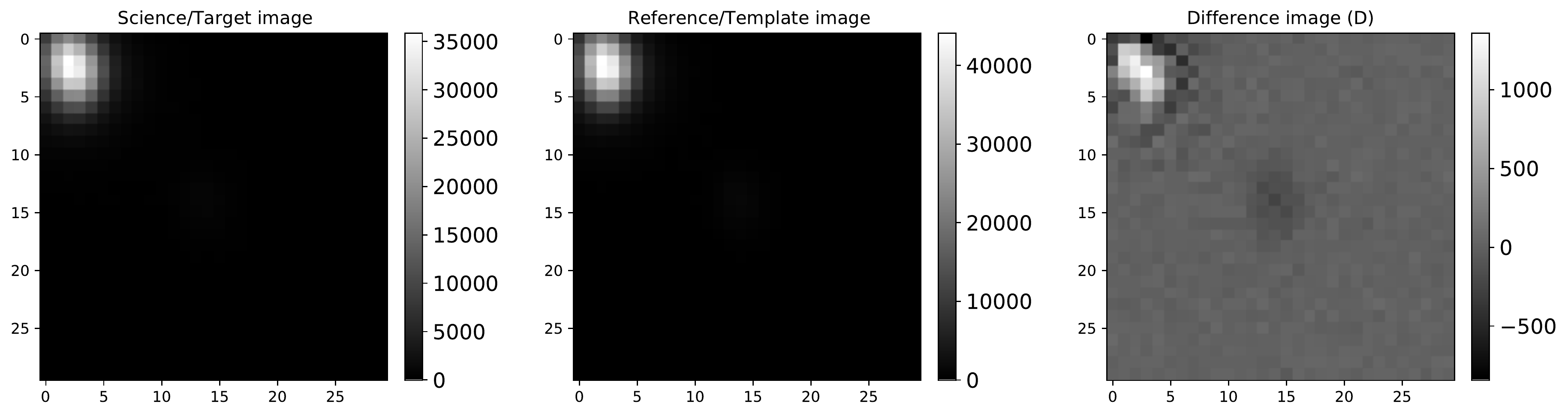}
    \caption{The false negatives obtained with \texttt{MeerCRAB3}, trained with \textit{T9} labelling and $NRD$ as input. These candidates have been classified as real by vetters. However, the best model misclassified them as bogus candidates. The first two rows illustrate two real candidates with characteristics of an extended source with > 10 pixels extent. The large extent of the sources can be attributed to that fact that the seeing can be very poor.}
    \label{fig:misclassification_real_1}
\end{figure}

In this section, we focus on our best model, \texttt{MeerCRAB3} network with the $NRD$ configuration and \textbf{T9} as input. In Figure \ref{fig:misclassification_bogus_1} and Figure \ref{fig:misclassification_real_1}, we present all misclassifications it made.

In Figure \ref{fig:misclassification_bogus_1}, the top two rows show bogus candidates with the presence of elongated spikes but the CNN found high pixel values at the centre, thus got confused and classified them as a real. The last row shows bogus candidates caused by blinking tail light of planes which can be recognized by the stripe. However, the \texttt{MeerCRAB} models seemingly recognized it as real due to the presence of a strong point source at the centre. However, we find that the decision made by \texttt{MeerCRAB} fulfils our requirements as we do not feed information related to elongated spikes and tail-light-trails when training the ML algorithm. These are not commonly occurring bogus events at present but may be added to our training set in the future and it will enable \texttt{MeerCRAB} to identify them.

Figure \ref{fig:misclassification_real_1} shows real candidates being misclassified as bogus. The top two rows are likely to be real point sources because the Seeing can be very poor, thus leading to the large extent of > 10 pixels. The fact that the source in the top row is oriented differently in the science/new and reference image suggests that it is not a Galaxy. However, these misclassifications are expected as these characteristics of real were not provided when training the network.

\section{Conclusions}\label{sec:conclusions}

A deep learning framework, \texttt{MeerCRAB}, integrated in the MeerLICHT facility is a step forward in the automation and improvement of the transient vetting process. In practice, by using \texttt{MeerCRAB} we can significantly reduce the number of missed transients per night and this may have a great impact on detecting and classifying the unknown unknowns of our universe.

In this paper, we detailed the process of developing \texttt{MeerCRAB}. To be able to train a deep neural network, we construct a large, high-quality labelled and representative dataset. To do so, we developed a vetting guidelines for vetters and taught them how real or bogus candidates in the MeerLICHT data appear. Then, a sample of 5000 candidates were provided to 10 vetters for labelling. Based on the vetters labels, we applied two methods to assign the final labelling to each candidate: (i) the \textit{thresholding} method (\textbf{T8}, \textbf{T9} \& \textbf{T10}) and (ii) latent class model $\textrm{L}_{lcm}$. At \textbf{T9}, a source is labelled as real if \textit{atleast 9} out of the 10 vetters labelled it as real or vice-versa. We found that \textbf{T9} is a good threshold criteria to be used since we have enough samples for training and testing the models, hence providing high quality labelled data. We found that going lower than this (i.e. \textbf{T7}, \textbf{T8}) or using $\textrm{L}_{lcm}$, added noisy labels. When used to train the network, such data decreased the performance of the models.

Moreover, we demonstrated that by increasing the depth of the network, (\texttt{MeerCRAB1} to \texttt{MeerCRAB3}), the performance of the model increases as well. McNemar's statistical test showed that \texttt{MeerCRAB3} performs better than \texttt{MeerCRAB1} and \texttt{MeerCRAB2}. In addition, we used a combination of input images (new ($N$), reference ($R$), difference ($D$), significance ($S$)) as input to the three networks. We found that with only $NR$, we obtained competitive results. We also observed that adding the difference and significance images improves network performance. 

In summary the best performing model has the following configuration: \textbf{T9} with \texttt{MeerCRAB3} having $NRD$ as input. This model yields an accuracy of 99.5 \% and MCC value of 0.989. This performance achieves an acceptable false positive and false negative rate for the real-time MeerLICHT transient detection pipeline requirements.

\texttt{MeerCRAB} is a crucial component of the MeerLICHT project which aims to detect and identify transient and variable sources. With the streaming data coming from MeerLICHT, the vast majority of astrophysical data are not only challenging to store, but also to classify efficiently and effectively. Therefore, \texttt{MeerCRAB} will enable the rapid identification of promising astrophysical sources in timely-manner. In addition, \texttt{MeerCRAB} can be adapted to be a system that disentangles interesting objects from a noisy background. We have already implemented similar models in radio astronomy that distinguish Single Pulses from Radio Frequency Interference for the MeerKAT telescope (\texttt{FRBID}\footnote{See \href{https://github.com/Zafiirah13/multi_input_frbid}{https://github.com/Zafiirah13/multi\_input\_frbid} and \href{https://github.com/Zafiirah13/FRBID}{https://github.com/Zafiirah13/FRBID} }: Fast Radio Burst Intelligent Distinguisher). \texttt{MeerCRAB} is a flexible software system, thus we were able to easily modify it to integrate different images as its inputs and as result, achieved high levels of performance when using it for radio astronomy images. Given the performance of \texttt{MeerCRAB} on both optical and radio image sources in astronomy, the method may have utility for those working in related areas.

\begin{table}
  \centering
  \caption{The results for various labelling methods are presented in terms of precision, recall, accuracy and MCC values using \textit{NRD} as input to the three models.}
  \label{tab:results}
  \begin{tabular}[0.5\textwidth]{ccccc} % four columns, alignment for each
    \hline
    \textbf{Methods of labelling} & \textbf{Precision} & \textbf{Recall} & \textbf{Accuracy} & \textbf{MCC}\\
    \hline
    \hline
    \multicolumn{5}{c}{\texttt{MeerCRAB1}}\\
    \hline
    $\textrm{L}_{lcm}$ & 0.96  & 0.96 & 0.960 & 0.920 \\
    T8  & 0.98 & 0.98 & 0.980 & 0.958\\
    T9  & 0.98 & 0.98 & 0.979 & 0.958\\
    T10   & 0.99 & 0.99 & 0.991 & 0.983\\
    \hline
    \multicolumn{5}{c}{\texttt{MeerCRAB2}}\\
    \hline
    $\textrm{L}_{lcm}$ & 0.97  & 0.97 & 0.967 & 0.936\\
    T8  & 0.99 & 0.98 & 0.977 & 0.953\\
    T9  & 0.99 & 0.99 & 0.986 & 0.973\\
    T10   & 0.99 & 0.99 & 0.994 & 0.988\\
    \hline
    \multicolumn{5}{c}{\texttt{MeerCRAB3}}\\
    \hline
    $\textrm{L}_{lcm}$ & 0.97 & 0.97 & 0.968 & 0.936\\
    T8  & 0.99 & 0.99 & 0.988 & 0.976\\
    \textbf{T9}  & \textbf{0.99} & \textbf{0.99} & \textbf{0.995} & \textbf{0.989}\\
    T10   & 1.00 & 1.00 & 0.998 & 0.995\\
    \hline
  \end{tabular}
\end{table}

\begin{table}
	\centering
	\caption{The results with Threshold 8 (\textbf{T8}) are presented in terms of precision, recall, accuracy and MCC values using various combinations of input images (new ($N$), reference ($R$), difference ($D$), and significance ($S$)) to the three models.}
	\label{tab:threshold_8}
	\begin{tabular}{ccccc} % four columns, alignment for each
		\hline
		Number of Images & Precision & Recall & Accuracy & MCC\\
		\hline
		\hline
		\multicolumn{5}{c}{\texttt{MeerCRAB1}}\\
		\hline
		NRDS & 0.98 & 0.98 & 0.980 & 0.958\\
		NRD  & 0.98 & 0.98 & 0.977 & 0.954\\
		NRS  & 0.97 & 0.97 & 0.971 & 0.942\\
		NR   & 0.96 & 0.96 & 0.962 & 0.923\\
		D    & 0.88 & 0.88 & 0.877 & 0.762\\
		S    & 0.83 & 0.83 & 0.831 & 0.662\\
		\hline
		\multicolumn{5}{c}{\texttt{MeerCRAB2}}\\
		\hline
		NRDS & 0.97 & 0.97 & 0.975 & 0.948\\
		NRD  & 0.98 & 0.98 & 0.977 & 0.953\\
		NRS  & 0.99 & 0.99 & 0.986 & 0.973\\
		NR   & 0.99 & 0.99 & 0.987 & 0.975\\
		D    & 0.91 & 0.91 & 0.910 & 0.825\\
		S    & 0.89 & 0.89 & 0.887 & 0.777\\
		\hline
		\multicolumn{5}{c}{\texttt{MeerCRAB3}}\\
		\hline
		NRDS & 0.98 & 0.98 & 0.983 & 0.966\\
		NRD  & \textbf{0.99} & \textbf{0.99} & \textbf{0.988} & \textbf{0.976}\\
		NRS  & 0.98 & 0.98 & 0.983 & 0.968\\
		NR   & 0.98 & 0.98 & 0.981 & 0.963\\
		D    & 0.92 & 0.92 & 0.921 & 0.843\\
		S    & 0.89 & 0.89 & 0.891 & 0.786\\
		\hline

	\end{tabular}
\end{table}

\begin{table}
	\centering
	\caption{The results with Threshold 9 (\textbf{T9}) are presented in terms of precision, recall, accuracy and MCC values using various combinations of input images (new ($N$), reference ($R$), difference ($D$), and significance ($S$)) to the three models.}
	\label{tab:threshold_9}
	\begin{tabular}{ccccc} % four columns, alignment for each
		\hline
		Number of Images & Precision & Recall & Accuracy & MCC\\
		\hline
		\hline
		\multicolumn{5}{c}{\texttt{MeerCRAB1}}\\
		\hline
		NRDS & 0.98 & 0.98 & 0.980 & 0.960\\
		NRD  & 0.98 & 0.98 & 0.979 & 0.958\\
		NRS  & 0.98 & 0.98 & 0.978 & 0.956\\
		NR   & 0.97 & 0.97 & 0.972 & 0.946\\
		D    & 0.86 & 0.83 & 0.823 & 0.690\\
		S    & 0.86 & 0.85 & 0.853 & 0.708\\
		\hline
		\multicolumn{5}{c}{\texttt{MeerCRAB2}}\\
		\hline
		NRDS & 0.99 & 0.99 & 0.988 & 0.976\\
		NRD  & 0.99 & 0.99 & 0.986 & 0.973\\
		NRS  & 0.99 & 0.99 & 0.986 & 0.973\\
		NR   & 0.99 & 0.99 & 0.989 & 0.978\\
		D    & 0.91 & 0.91 & 0.912 & 0.827\\
		S    & 0.89 & 0.87 & 0.865 & 0.751\\
		\hline
		\multicolumn{5}{c}{\texttt{MeerCRAB3}}\\
		\hline
		NRDS & 0.99 & 0.99 & 0.987 & 0.974\\
		NRD  & \textbf{0.99} & \textbf{0.99} & \textbf{0.995} & \textbf{0.989}\\
		NRS  & 0.99 & 0.99 & 0.990 & 0.980\\
		NR   & 0.99 & 0.99 & 0.989 & 0.978\\
		D    & 0.93 & 0.93 & 0.931 & 0.863\\
		S    & 0.89 & 0.87 & 0.868 & 0.760\\
		\hline

	\end{tabular}
\end{table}

\begin{table}
	\centering
	\caption{The results with Threshold 10 (\textbf{T10}) are presented in terms of precision, recall, accuracy and MCC values using various combinations of input images (new ($N$), reference ($R$), difference ($D$), and significance ($S$)) to the three models.}
	\label{tab:threshold_10}
	\begin{tabular}{ccccc} % four columns, alignment for each
		\hline
		Number of Images & Precision & Recall & Accuracy & MCC\\
		\hline
		\hline
		\multicolumn{5}{c}{\texttt{MeerCRAB1}}\\
		\hline
		NRDS & 1.00 & 1.00 & 0.995 & 0.990\\
		NRD  & 0.99 & 0.99 & 0.991 & 0.983\\
		NRS  & 0.98 & 0.98 & 0.985 & 0.970\\
		NR   & 0.99 & 0.99 & 0.986 & 0.973\\
		D    & 0.92 & 0.92 & 0.920 & 0.841\\
		S    & 0.93 & 0.93 & 0.934 & 0.868\\
		\hline
		\multicolumn{5}{c}{\texttt{MeerCRAB2}}\\
		\hline
		NRDS & 1.00 & 1.00 & 0.995 & 0.990\\
		NRD  & 0.99 & 0.99 & 0.994 & 0.988\\
		NRS  & 0.98 & 0.98 & 0.984 & 0.968\\
		NR   & 0.99 & 0.99 & 0.992 & 0.985\\
		D    & 0.94 & 0.94 & 0.944 & 0.888\\
		S    & 0.94 & 0.94 & 0.940 & 0.881\\
		\hline
		\multicolumn{5}{c}{\texttt{MeerCRAB3}}\\
		\hline
		NRDS & 0.99 & 0.99 & 0.990 & 0.980\\
		NRD  & \textbf{1.00} & \textbf{1.00} & \textbf{0.998} & \textbf{0.995}\\
		NRS  & 0.99 & 0.99 & 0.994 & 0.988\\
		NR   & 0.99 & 0.99 & 0.985 & 0.970\\
		D    & 0.94 & 0.94 & 0.939 & 0.878\\
		S    & 0.94 & 0.94 & 0.943 & 0.886\\
		\hline

	\end{tabular}
\end{table}

\begin{table}
	\centering
	\caption{McNemar's Test using \textit{T9} and \textit{NRD} results for model selection. \texttt{MeerCRAB3} has the best performance than \texttt{MeerCRAB1} and \texttt{MeerCRAB2} as the p-value is less than 0.05.}
	\label{tab:McNemar_test}
	\begin{tabular}{cc} % four columns, alignment for each
		\hline
		Models & P-value \\
		\hline
		\texttt{MeerCRAB1} vs \texttt{MeerCRAB2} & 0.13400\\
		\texttt{MeerCRAB2} vs \texttt{MeerCRAB3} & 0.00390\\
		\texttt{MeerCRAB1} vs \texttt{MeerCRAB3} & 0.00008\\
		\hline

	\end{tabular}
\end{table}

\section{Declarations}

\section*{Funding}
ZH acknowledges support from the UK Newton Fund as part of the Development in Africa with Radio Astronomy (DARA) Big Data project delivered via the Science \& Technology Facilities Council (STFC). BWS acknowledges funding from the European Research Council (ERC) under the European Union\textquotesingle s Horizon 2020 research and innovation programme (grant agreement No. 694745). PJG and SDW are supported by NRF SARChI Grant 111692. 

\section*{Conflicts of interest/Competing interests}
Not applicable

\section*{Availability of data and material}
Data will be available upon request.

\section*{Code availability}
\texttt{MeerCRAB} code and pre-trained models are available on Github at \href{https://github.com/Zafiirah13/meercrab}{https://github.com/Zafiirah13/meercrab} and on Zenodo at \href{https://doi.org/10.5281/zenodo.4049943}{https://doi.org/10.5281/zenodo.4049943}. 

\section*{Acknowledgements}
We thank the referee for useful comments and suggestions for improving this paper. We would like to thank the people who gave up their time to do the vetting of the sample: Laura Driessen, Naomi titus, Mark Beijer, Nadia Blagorodnova, Joris Kersten, David Modiano and Roque Ruiz Carmona, without whose effort this work would not have been possible. We would like to also thank Arrykrishna Mootoovaloo and Fabian Gieseke for useful discussion. The MeerLICHT consortium is a partnership between Radboud University, the University of Cape Town, the Netherlands Organisation for Scientific Research (NWO), the South African Astronomical Observatory (SAAO), the University of Oxford, the University of Manchester and the University of Amsterdam, in association with and, partly supported by, the South African Radio Astronomy Observatory (SARAO), the European Research Council and the Netherlands Research School for Astronomy (NOVA).

\bibliographystyle{aa_url} % use for links in references

\bibliography{combined}

\end{document}